%% file: chirpotle.tex
\documentclass[authorversion,sigconf]{acmart}
\pdfoutput=1
\usepackage[utf8]{inputenc}

\usepackage{flushend}
\usepackage{balance}

\usepackage{tikz}
\usepackage{xifthen}
\usepackage[nolist]{acronym}
\usepackage{csvsimple}
\usepackage{datatool}
\usepackage[binary-units]{siunitx}
\DeclareSIUnit\dBm{\dB\text{m}}

\input{inc/tikzdefines}

\copyrightyear{2020}
\acmYear{2020}
\setcopyright{acmlicensed}\acmConference[WiSec '20]{13th ACM Conference on Security and Privacy in Wireless and Mobile Networks}{July 8--10, 2020}{Linz (Virtual Event), Austria}
\acmBooktitle{13th ACM Conference on Security and Privacy in Wireless and Mobile Networks (WiSec '20), July 8--10, 2020, Linz (Virtual Event), Austria}
\acmPrice{15.00}
\acmDOI{10.1145/3395351.3399423}
\acmISBN{978-1-4503-8006-5/20/07}

\newcommand{\frameworkname}{ChirpOTLE}

\acmSubmissionID{wisec20-p121}

\begin{document}
\begin{filecontents*}{data/adr_trig.csv}
whtype,datarate,direct,indirect,failed
1,0,16,44,0
1,1,26,33,1
1,2,23,37,0
1,3,19,41,0
2,0,0,54,6
2,1,0,56,4
2,2,60,0,0
2,3,58,2,0
\end{filecontents*}

\begin{filecontents*}{data/timing_info_rx2.csv}
datarate,proc1_mean,proc2_mean,uplink_mean,downlink_mean,rxdelay1_mean,rxdelay2_mean,proc1_std,proc2_std,uplink_std,downlink_std,rxdelay1_std,rxdelay2_std,proc1_n,proc2_n,uplink_n,downlink_n,rxdelay1_n,rxdelay2_n
0,183.60658011049725,0,1161.4083535911602,0,0,0,29.85699086447178,0,31.59111893465443,0,0,0,181,0,181,0,0,0
1,167.85022900763357,121.949,659.456,659.456,205.035,1781.518,27.378171401521087,0.0,0.0,0.0,0.0,0.0,131,1,131,1,1,1
2,166.09786666666668,243.81367924528303,288.768,329.728,999.9188679245283,2028.0234716981133,31.071083907424637,30.230301478828988,0.0,0.0,0.004576551127292158,0.005045467884449896,60,53,60,53,53,53
3,159.49923333333334,515.7855111111111,164.864,164.864,1000.1167555555555,2009.3478666666667,26.40617270852068,87.0397909333228,0.0,0.0,0.00471477844680629,86.68782522710119,60,45,60,45,45,45
5,143.2476402569593,792.7401400651466,46.336,46.336,1005.0944690553746,2023.8243127035832,33.70605231521075,62.44153722545675,0.0,0.0,0.005110650572143364,51.92141846502397,467,307,467,307,307,307
\end{filecontents*}

\input{inc/acronyms}

\title{\frameworkname{}: A Framework for Practical LoRaWAN Security Evaluation}

\author{Frank Hessel}
\authornote{Both authors contributed equally to this research.}
\orcid{0000-0001-8158-3030}
\affiliation{%
	\institution{Secure Mobile Networking Lab\\
	Department of Computer Science\\
	TU Darmstadt, Germany}
}
\email{fhessel@seemoo.de}

\author{Lars Almon}
\authornotemark[1]
\orcid{0000-0003-1296-2920}
\affiliation{%
	\institution{Secure Mobile Networking Lab\\
	Department of Computer Science\\
	TU Darmstadt, Germany}
}
\email{lalmon@seemoo.de}

\author{Flor Álvarez}
\orcid{0000-0002-0584-6406}
\affiliation{%
	\institution{Secure Mobile Networking Lab\\
	Department of Computer Science\\
	TU Darmstadt, Germany}
}
\email{falvarez@seemoo.de}


\begin{abstract}
\input{inc/abstract}
\end{abstract}



\maketitle

\newpage

\input{inc/intro}

\input{inc/background}

\input{inc/related-work}

\input{inc/framework}

\input{inc/attacks}

\input{inc/setup}

\input{inc/evaluation}

\input{inc/conclusion}

\begin{acks}
This work has been co-funded by the LOEWE initiative (Hessen State Ministry for Higher Education, Research and the Arts, Germany) within the emergenCITY centre and by the German Federal Ministry of Education and Research and the Hessen State Ministry for Higher Education, Research and the Arts within their joint support of the National Research Center for Applied Cybersecurity ATHENE.
\end{acks}

\balance
\bibliographystyle{ACM-Reference-Format}
\bibliography{chirpotle}

\end{document}

%% file: inc/tikzdefines.tex
\graphicspath{{./img/}}

\usetikzlibrary{arrows.meta,decorations.pathreplacing,angles,quotes,patterns}

\usetikzlibrary[positioning]

\makeatletter
\def\pgfplots@stacked@diff{}
\makeatother

\definecolor{verylightgray}{gray}{0.85}
\definecolor{lightgray}{gray}{0.75}
\definecolor{darkgray}{gray}{0.5}

\definecolor{inkscape-royalblue}{RGB}{65,105,225}
\definecolor{inkscape-firebrick}{RGB}{178,34,34}
\definecolor{inkscape-forestgreen}{RGB}{85,155,85}
\definecolor{inkscape-khaki}{RGB}{240,230,140}
\definecolor{inkscape-lightsteelblue}{RGB}{176,196,222}

\usepackage{pgfplots}
\usepgfplotslibrary{groupplots}
\usepgfplotslibrary{colorbrewer}

\pgfplotsset{
	discard if not/.style 2 args={
		x filter/.code={
			\edef\tempa{\thisrow{#1}}
			\edef\tempb{#2}
			\ifx\tempa\tempb
			\else
			
			\fi
		}
	}
}

\tikzset{annotation/.style={scale=0.85,fill=white,inner sep=1,outer sep=3}}
\tikzset{txarrow/.style={-{Latex[scale=1.5]},dashed}}

\newcommand{\tikzdrawframe}[5]{
	\node [draw,fill=white,rectangle,minimum height=4*#5,minimum width=3] (#4_preamble) at (#1,#2) {};
	\node [draw,fill=#3,rectangle,minimum height=10*#5,minimum width=3] (#4_payload) [below=of #4_preamble] {};
}

\newcommand{\tikzdrawwindow}[3]{
	\node [draw,dotted,fill=white,rectangle,minimum height=4,minimum width=10] (#3) at (#1,#2) {};
}

\DTLloaddb[noheader=false]{trigfile}{data/adr_trig.csv}
\DTLloaddb[noheader=false]{rx2timingfile}{data/timing_info_rx2.csv}

%% file: inc/acronyms.tex
\begin{acronym}
\acro{abp}[ABP]{activation by personalization}
\acro{adr}[ADR]{adaptive data rate}
\acro{as}[AS]{application server}
\acro{css}[CSS]{chirp-spread spectrum}
\acro{dos}[DoS]{denial-of-service}
\acro{dr}[DR]{data rate}
\acro{ed}[ED]{end device}
\acro{fcnt}[FCnt]{frame counter}
\acro{js}[JS]{join server}
\acro{gw}[GW]{gateway}
\acro{ism}[ISM]{industrial, scientific and medical}
\acro{lpwan}[LPWAN]{low-power wide-area network}
\acro{mac}[MAC]{medium access control}
\acro{mcu}[MCU]{microcontroller unit}
\acro{mic}[MIC]{message integrity code}
\acro{ns}[NS]{network server}
\acro{otaa}[OTAA]{over-the-air activation}
\acro{repl}[REPL]{read-eval-print loop}
\acro{rpc}[RPC]{remote procedure call}
\acro{rssi}[RSSI]{received signal strength indication}
\acro{sf}[SF]{spreading factor}
\acro{snr}[SNR]{signal-to-noise ratio}
\acro{tp}[TP]{transmission power}
\acro{ttn}[TTN]{The Things Network}
\acro{unb}[UNB]{ultra-narrow band}
\end{acronym}

%% file: inc/abstract.tex
Low-power wide-area networks (LPWANs) are becoming an integral part of the Internet of Things.
As a consequence, businesses, administration, and, subsequently, society itself depend on the reliability and availability of these communication networks.

Released in 2015, LoRaWAN gained popularity and attracted the focus of security research,
revealing a number of vulnerabilities.
This lead to the revised LoRaWAN~1.1 specification in late 2017.
Most of previous work focused on simulation and theoretical approaches.
Interoperability and the variety of implementations complicate the risk assessment for a specific LoRaWAN network.

In this paper, we address these issues by introducing \frameworkname{}, a LoRa and LoRaWAN security evaluation framework suitable for rapid iteration and testing of attacks in testbeds and assessing the security of real-world networks.
We demonstrate the potential of our framework by verifying the applicability of a novel denial-of-service attack targeting the adaptive data rate mechanism in a testbed using common off-the-shelf hardware.
Furthermore, we show the feasibility of the Class~B beacon spoofing attack, which has not been demonstrated in practice before.

%% file: inc/intro.tex
\section{Introduction}
\label{sec:intro}

Modern society increasingly relies on connected smart infrastructure.
From the Internet of Things to smart cities and cyber-physical systems, connectivity becomes the key enabler. 
Low-power wide-area networks \acused{lpwan} (\acp{lpwan}) are gaining more and more momentum to support this transition.
With technologies like Sigfox, NB-IoT, LTE-M, LoRaWAN and others competing in this field, the question for their security arises and has moved into focus of the scientific community. A thorough analysis of their security properties, potential threats, and countermeasures is fundamental for the resilient and reliable operation of connected infrastructure.

We want to foster the research on \ac{lpwan} security with a focus on LoRaWAN.
This protocol stands out for its open operator model and open-source software stacks like ChirpStack \cite{chirpstack-ns} and The Things Stack \cite{ttn}, which enable community-based networks.

While this accessibility of the technology clearly contributed to LoRaWAN's popularity, interoperability and implementation-specific details come with an additional class of security issues.
Most currently known problems affect the LoRaWAN specification in version 1.0 and have been addressed in LoRaWAN 1.1 \cite{donmez2018security}.
Research on this topic relies mostly on theoretical discussion or simulation to confirm potential findings.
This focus and the heterogeneous environment of software versions and specifications raise the demand for practical assessment of the feasibility of attacks and countermeasures in LoRaWAN networks.

In this paper, we present \frameworkname{}, a novel LoRa and LoRaWAN security evaluation framework, to ease practical experimentation in testbeds and security assessments in real-world LoRaWAN networks.
\frameworkname{} orchestrates distributed off-the-shelf LoRa nodes and comes with the building blocks to rapidly find and validate potential vulnerabilities.

After giving background information on LoRaWAN in Section~\ref{sec:background} and presenting related work in Section~\ref{sec:related-work}, we make the following contributions:
\begin{itemize}
 \item First, we present an open-source security evaluation framework for LoRaWAN (Section~\ref{sec:framework}).\footnote{\href{https://github.com/seemoo-lab/chirpotle}{https://github.com/seemoo-lab/chirpotle}}
 \item Second, we introduce the concept of wormholes in LoRaWAN to propose a novel \ac{dos} attack exploiting the \ac{adr} mechanism (Section~\ref{sec:wormholes}).
 \item Third, we present the first experimental evaluation of the Class~B beacon spoofing attack (Section~\ref{sec:beacon-spoofing}).
 \item Finally, we discuss results and possible countermeasures (Sections~\ref{sec:setup} and \ref{sec:evaluation}). 
\end{itemize}

\noindent We conclude and summarize our work in Section~\ref{sec:conclusion}.

%% file: inc/background.tex
\section{Background on LoRaWAN}
\label{sec:background}

LoRaWAN is a specification for an infrastructure to connect sensor nodes with centrally managed applications and for \ac{mac} using LoRa as physical layer technology.

A LoRaWAN network consists of \acp{ed} which communicate with a central \ac{ns}. Messages are forwarded through a LoRa link between the \ac{ed} and one or more \acp{gw} followed by a backing network connection to the \ac{ns}.
Application data is processed on \acp{as} after being forwarded to them by the \ac{ns}.
In advance to all other communication, \acp{ed} and the \ac{ns} establish a device session either interactively using a mechanism called \ac{otaa} or statically with \ac{abp}.

In this paper, we focus on the wireless LoRa link.
By default, all communication is initiated by the \ac{ed}, which is called Class~A operation.
The \ac{ed} transmits an uplink message and then waits for a period $d_{rx_1}$ before opening the first receive window called ``rx1''.
After an additional delay\footnote{In contrast to the use of RECEIVE\_DELAY2 in the LoRaWAN specification, we define $d_{rx_2}$ as the delay \emph{between} receive windows.} of $d_{rx_2}$, a second receive window (``rx2'') is opened.
$d_{rx_1}$ can be configured in a range between \SI{1}{\second} and \SI{15}{\second}, while $d_{rx_2}$ is a fixed value of \SI{1}{\second} \cite[Section 5.8]{lora2017specs11}.
We use the term \emph{transaction} to refer to an uplink message and the optional downlink response in a corresponding receive window.

The payload of all data messages is encrypted, a \ac{mic} is appended to messages to protect their integrity and authenticity, and distinct \acp{fcnt} for each communication direction prevent replay attacks.

LoRaWAN manages its \ac{mac} layer by the exchange of so-called MAC commands.
These MAC commands can either be sent in the payload field of a message, if no application data is pending, or piggy-backed to a data message in a special \texttt{FOpts} field.

Region-specific differences of the specification allow operation on the \ac{ism} bands. 
We focus on the \texttt{EU868} region. We now introduce two relevant LoRaWAN features for our analysis: \ac{adr} and Class~B Operation.

\subsection{Adaptive Data Rate}
\label{sec:background-adr}

Due to the nature of a \ac{css} modulation like LoRa, transmission times grow quadratic with decreasing data rate.
The same payload can occupy the medium for milliseconds or seconds, which makes selecting an appropriate \ac{dr} crucial for performance in dense, large-scale networks, especially in constrained \ac{ism} bands.
LoRaWAN employs \ac{adr} to address this issue \cite[Section 4.3.1.1]{lora2017specs11}.
Its goal is to keep each \ac{ed} at a certain demodulation margin, which is the set screw between quality of service and efficient bandwidth allocation.

An \ac{ed} activates \ac{adr} by setting the \texttt{ADR} flag in its uplink messages.
The \ac{ns} then collects information on the reception quality for the \ac{ed} and once enough data is gathered, issues a \texttt{LinkADRReq} MAC command, which contains a target \ac{dr} and \ac{tp} for the \ac{ed}.
To cope with frame loss due to collisions, the command can also be used to limit the \ac{ed} to specific channels and to request redundant transmissions using the \texttt{nbTrans} field.
By accepting the request using the \texttt{LinkADRAns} MAC command, the \ac{ed} adjusts its \ac{dr} for further transmissions.

To assure connectivity, the \ac{ed} keeps track of the number of transactions since the last downlink message in a value called \texttt{ADR\_\allowbreak{}ACK\_\allowbreak{}CNT}.
If this value exceeds a configurable threshold called \texttt{ADR\_\allowbreak{}ACK\_\allowbreak{}LIMIT} (default: \num{64}), it sets the \texttt{ADRACKReq} flag in each uplink to encourage the \ac{ns} to schedule a downlink message.
Whenever the \ac{ed} receives a downlink message, it resets its \texttt{ADR\_\allowbreak{}ACK\_\allowbreak{}CNT} to \num{0}.
If \texttt{ADR\_\allowbreak{}ACK\_\allowbreak{}CNT}, however, exceeds \texttt{ADR\_\allowbreak{}ACK\_\allowbreak{}LIMIT}~+ \texttt{ADR\_\allowbreak{}ACK\_\allowbreak{}DELAY}, with the latter being another configurable parameter, the \ac{ed} starts to increase its \ac{tp} and \ac{dr} stepwise every \texttt{ADR\_\allowbreak{}ACK\_\allowbreak{}DELAY} transactions, until it eventually receives a response.

Summarized, the \ac{adr} mechanism is device-induced but network-controlled.
The LoRaWAN specification does not define the exact algorithm to use on the \ac{ns}, but Semtech has published a recommendation for an algorithm \cite{semtech2016adr}, that has found its way into LoRaWAN software like ChirpStack\footnote{cf. \url{https://github.com/brocaar/chirpstack-network-server/blob/v3.6.0/internal/adr/}}.
It collects server-side \ac{snr} measurements for up to \num{20} frames and uses their maximum to issue the \texttt{LinkADRReq} command.
The authors justify the usage of the maximum by interference being the main reason for packet loss.
Furthermore and in contrast to LoRaWAN~1.0 \cite[Section 4.3.1.1]{lora2018specs103}, LoRaWAN~1.1 explicates that \ac{adr} should be requested for stationary \acp{ed} even if the \ac{ns} signals its inability to appropriately estimate a good configuration.
We show how both of these decisions endanger the reliability of a network under attack.

\subsection{Class B Operation}
\label{sec:background-classb}

In Class~A, downlink traffic must follow uplink messages.
Applications with requirements on downlink latency can operate in Class~B, whereby \acp{ed} open additional receive windows.
Temporal synchronization is achieved by transmitting network-global beacons every \SI{128}{\second} from \acp{gw} based on a global, GPS-synchronized clock.

The \acp{ed} use beacon metadata (time of arrival), beacon payload (GPS timestamp), and device properties (device address) to calculate receive window offsets.
Including the address leads to pseudo-random offsets for each \ac{ed} and reduces collisions.

Since an \ac{ed} may be within reach of multiple \acp{gw} and to allow the coexistence of networks from different operators, all beacons are sent simultaneously to aim for constructive interference \cite[Section~15.2, Note~1]{lora2017specs11}.
As a direct consequence, the payload of beacons cannot undergo meaningful encryption nor authentication, since keys would need to be available for all network operators.

To enable Class~B, the \ac{ed} starts by searching and locking to beacons frames.
Once a stable lock exists, it transmits all further uplink messages with the \texttt{ClassB} flag set to notify the \ac{ns} about the class switch.
This process can be sped up by requesting the current time from the \ac{ns} using the \texttt{DeviceTimeReq} MAC command, allowing the \ac{ed} to estimate the next arrival of a beacon.
When the lock is lost, the \ac{ed} runs at least two hours of ``beacon-less operation'', in which it gradually widens all Class~B receive windows before eventually switching back to Class~A \cite[Section 5.12]{lora2017specs11}.

%% file: inc/related-work.tex
\section{Related Work}
\label{sec:related-work}

We present the relevant related work in three parts: Studies on \ac{adr} performance, on jamming and wormholes, and on beacon spoofing.

Most previous work on \ac{adr} covers performance aspects.
\citeauthor{li2018agile} give a simulation-based performance evaluation of Semtech's \ac{adr} algorithm with a focus on time to convergence \cite{li2018agile}.
Their results show that approaching convergence from high \acp{dr} is notably more time-consuming than from low \acp{dr}, meaning that over-optimizing the \ac{dr} causes a prolonged decline of \ac{ed} availability.

Alternative server-side algorithms for \ac{adr} have been proposed for a variety of optimization goals.
\citeauthor{bor2017lora} suggest using a probing-based mechanism to determine the best \ac{dr} for each device \cite{bor2017lora}.
Most other approaches, however, strive for global optimization by considering all \acp{ed} within a certain region.
\citeauthor{reynders2017power} propose a cell-based algorithm that aims for fairer packet error rates in the outer region of a cell based on node distance or path-loss estimation \cite{reynders2017power}.
\citeauthor{abdelfadeel2018fair} extend this work and suggest to use fairness regarding equal overall collision probability as a goal \cite{abdelfadeel2018fair}.
\citeauthor{cuomo2017explora} designate the greediness of the default \ac{adr} algorithm to use high \acp{dr} as a main cause for network congestion and also propose two alternative schemes.
They exploit the orthogonality of different \acp{sf} to equalize \ac{dr} usage either by the number of devices or by expected air time \cite{cuomo2017explora}.
To our best knowledge, none of the proposed algorithms considers the presence of intentionally manipulated \ac{snr} measurements.

Due to its \ac{css} modulation, jamming LoRa is most effective by exploiting coexistence issues with \ac{css} signals.
\citeauthor{goursaud2015dedicated} compare \ac{css} to \ac{unb} modulation and show that LoRa mainly interferes with other LoRa signals \cite{goursaud2015dedicated}.
They prove that interference between LoRa frames of different \acp{sf} is low while a signal of the same \ac{sf} has to be at least \SI{6}{\dB} stronger to demodulate it by benefiting from the capture effect.
\citeauthor{croce2018impact} underline these findings by simulations and experiments and also point out that orthogonality between \acp{sf} is limited \cite{croce2018impact}.

\citeauthor{aras2017exploring} use this knowledge to create a LoRa jammer based on off-the-shelf hardware which can perform triggered and payload-based reactive jamming \cite{aras2017exploring,aras2017selective}.
They also evaluate a unidirectional store-and-forward wormhole based on two nodes which handles frames of SF10 and above reliably.

\citeauthor{miller2016lora} mentions a beacon spoofing attack shortly after the publication of LoRaWAN 1.0.0, either to manipulate the beacon's payload or to tamper with the calculation of receive windows \cite{miller2016lora}.
Van Es et al. formally verify the possibility of injecting malicious beacons with modified time values to provoke calculating invalid receive window offsets on the \acp{ed} \cite{vanes2018denial}.
\citeauthor{yang2018security} also discuss the impact of modification to different beacon payload fields \cite{yang2018security}.
Contrary to their proposition, it is not possible to change the wakeup periodicity and drain \ac{ed}'s battery by changing the beacon payload.
\citeauthor{butun2018analysis} assess beacon spoofing to be relevant for LoRaWAN~1.1 \cite{butun2018analysis}.

%% file: inc/framework.tex
\section{Security Evaluation Framework}
\label{sec:framework}

LoRaWAN is a distributed and heterogeneous system built on an infrastructure consisting of \acp{ns}, \acp{as}, and \acp{gw}.
The different protocols and the network infrastructure connecting all of these components provide many options for conducting a security evaluation, but also demand for a specific attacker model when it comes to security evaluation.

\subsection{Attacker and Threat Model}
\label{sec:model}

We limit our scope to the wireless link between \acp{ed} and \acp{gw}.
This decision comes with the least preconditions in the attacker model, as the wireless communication channel is, by its nature, accessible to everyone within proximity.

Furthermore, we limit the attacker to use only inexpensive off-the-shelf LoRa hardware to emphasize the low requirements for reproducing the attacks.
Using only LoRa transceivers intended for use in \acp{ed}, however, comes with challenges:
Chips like Semtech's SX127x series can only handle one channel and \ac{dr} at a time.

We assume the attacker to be able to transmit and receive arbitrary LoRa frames on a given channel.
The location of devices, network configuration, and other publicly observable metadata like uplink periodicity of \acp{ed} is assumed to be known by the attacker.
However, the attacker is not in possession of any cryptographic keys or other data that is marked as protected in the LoRaWAN specification, nor she is able to break cryptographic primitives.

\subsection{Framework Design and Architecture}
\label{sec:framework-architecture}

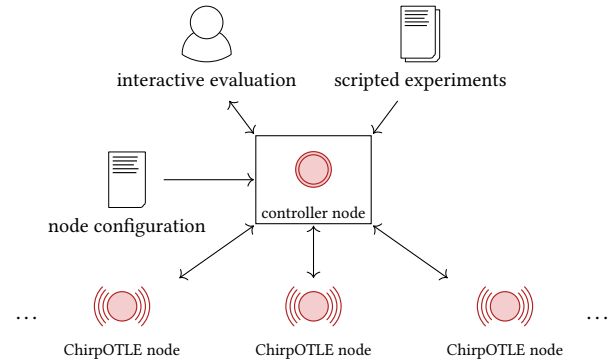
\begin{figure}[t]
	\centering
	\input{img/framework-architecture}
	\Description{A diagram showing the ``controller node´´ in the center. From the top, two icons labeled ``interactive evaluation´´ and ``scripted experiments´´ are connected to the controller. From the side, a ``node configuration´´ is added to the controller. At the bottom, multiple ``field nodes´´ are shown, each also connected to the central controller node.}
	\caption{Architecture of the \frameworkname{} framework}
	\label{fig:framework-architecture}
\end{figure}

Based on this attacker model, we design and implement \frameworkname{}, our LoRa and LoRaWAN security evaluation framework.
The main challenge is the physical layer interaction within the coverage area of the network under test.
Furthermore, attacks involving wormholes require forwarding frames between different locations, while some decisions are made and actions are run locally on the node, due to strict timing constraints.
Selective jamming, for instance, needs quick decisions for or against jamming a frame while it is still in transmission.
From these requirements, we identify two main components of the framework:
First, a set of LoRa field nodes with real-time support and a network interface for configuration and out-of-band communication.
Second, a controller that orchestrates their actions in complex interaction patterns.

Figure~\ref{fig:framework-architecture} gives a high-level overview of the resulting architecture.
Using a flexible, file-based node configuration allows fast adaption to given network topology and available hardware.
With Python as a high-level language on the controller node, we can use its built-in REPL interface for interactive assessment of vulnerabilities and evaluation of new attacks.
This environment inherently allows running scripts to generate quantitative experimental results without manual intervention.

To fulfill the real-time requirements and to address a great variety of off-the-shelf hardware, we implement a low-level \emph{companion application} based on RIOT \cite{baccelli2018riot}, an operating system for \acp{mcu}.
This application connects to SX127x-compatible LoRa radios through SPI and provides the host with a serial interface for higher-level commands like switching the channel or activating a jammer.

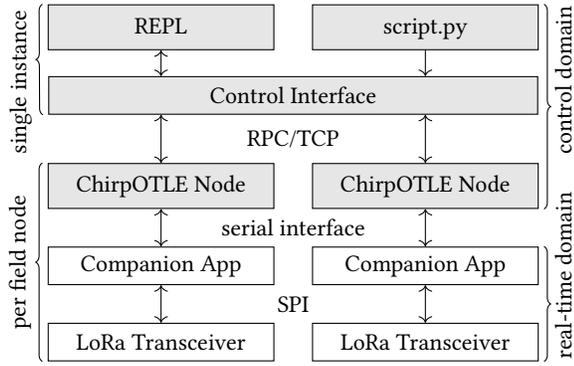
\begin{figure}[t]
	\centering
	\input{img/framework-control-flow}
	\Description{The diagram shows the control flow between different components of the frame work. The central entity is the ``control interface''. Above the control interface, there are the REPL, which can communicate with the interface interactively, and label called ``script.py'' which serves input to the control interface unidirectionally. All three nodes together are labled with ``single instance''. Below the control interface, two equal stacks are depicted. The controller reaches the next layer, called ``LoRa Node'' via RPC and TCP. From there, a serial interface is used to reach the ``companion application''. Finally, the ``LoRa transceiver'' is reached via SPI. The last three blocks are labeled ``per field node''. On the right side, the diagram also distinguished between the control domain, which contains the REPL, the script, the control interface, and the LoRa nodes, and the real-time domain, which contains the companion application and the LoRa transceiver.}
	\caption{Control flow in the security evaluation framework}
	\label{fig:framework-control-flow}
\end{figure}

To combine both, the low-level \ac{mcu} application and the high-level scripting environment for experiment design, we use the TPy framework \cite{steinmetzer2018tpy}.
It simplifies deployment and control in distributed network experiments.
Together with the node configuration, its \ac{rpc} abstraction layer hides communication details between controller and \frameworkname{} nodes in the field, as shown in Figure~\ref{fig:framework-control-flow}.
One or more MCUs executing the companion application can be hooked up to a lightweight \ac{rpc} host, for example, a Raspberry Pi, that handles network communication and translates \ac{rpc} calls into commands for the serial interface of the \ac{mcu}.

%% file: img/framework-architecture.tex
\begin{tikzpicture}[
	node distance=0.2cm
]
\tikzset{
	fieldnode/.style = {rectangle,minimum width=2cm, inner sep=0, outer sep=0},
	controller/.style = {rectangle,minimum width=2cm, inner sep=3, outer sep=1,draw},
}

\node[fieldnode] (fieldnode0) [align=center,scale=0.7] {\includegraphics[scale=0.714]{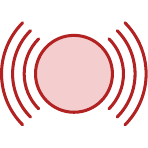}\\\frameworkname{} node};

\node[fieldnode] (fieldnode1) [right=1cm of fieldnode0,align=center,scale=0.7] {\includegraphics[scale=0.714]{img/icon-attacker-node.pdf}\\\frameworkname{} node};

\node[fieldnode] (fieldnode2) [right=1cm of fieldnode1,align=center,scale=0.7] {\includegraphics[scale=0.714]{img/icon-attacker-node.pdf}\\\frameworkname{} node};

\node (pseudofieldnode0) [left=of fieldnode0] {\dots};
\node (pseudofieldnode1) [right=of fieldnode2] {\dots};

\node[controller] (controller) [above=.7cm of fieldnode1,align=center,scale=0.7] {\includegraphics[scale=0.714]{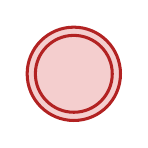}\\controller node};

\draw [<->] (controller.south west) -- (fieldnode0);
\draw [<->] (controller.south) -- (fieldnode1.north);
\draw [<->] (controller.south east) -- (fieldnode2);

\node (configgfx) [left=1.2cm of controller] {\includegraphics[scale=0.5]{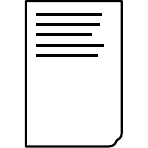}};
\node [annotation] (configlbl) [below=-0.1cm of configgfx] {node configuration};
\draw [->] (configgfx) -- (controller.west);

\node (interactivegfx) [above left=of controller,yshift=0.7cm] {\includegraphics[scale=0.5]{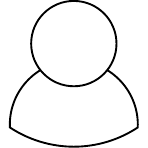}};
\node [annotation] (interactivelbl) [below=-0.1cm of interactivegfx] {interactive evaluation};
\draw [<->, shorten <=.5cm] (interactivegfx.south) -- (controller.north west);

\node (scriptsgfx) [above right=of controller,yshift=0.7cm] {\includegraphics[scale=0.5]{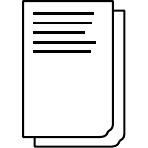}};
\node [annotation] (scriptslbl) [below=-0.1cm of scriptsgfx] {scripted experiments};
\draw [->, shorten <=.5cm] (scriptsgfx.south) -- (controller.north east);

\end{tikzpicture}

%% file: img/framework-control-flow.tex
\begin{tikzpicture}
\tikzset{
	block/.style={rectangle,minimum width=3cm,minimum height=.5cm},
	ctrlblock/.style={block,draw=black,fill=gray!20!white},
	rtblock/.style={block,draw=black},
	dblsize/.style={minimum width=6.5cm},
}

\node [ctrlblock] (repl) {\strut{}REPL};
\node [ctrlblock] (script) [right=.5cm of repl] {\strut{}script.py};

\path (repl) -- node [ctrlblock, dblsize] (controlinterface) [below=.65cm] {Control Interface} (script);
\draw [<->] (repl) -- (repl|-controlinterface.north);
\draw [->] (script) -- (script|-controlinterface.north);

\node [ctrlblock] (loranode0) [below=1.5cm of repl] {\strut{}\frameworkname{} Node};
\node [ctrlblock] (loranode1) [below=1.5cm of script] {\strut{}\frameworkname{} Node};
\draw [<->] (loranode0|-controlinterface.south) -- node [right,align=center,minimum width=3.5cm] {RPC/TCP} (loranode0);
\draw [<->] (loranode1|-controlinterface.south) -- (loranode1);

\node[rtblock] (companion0) [below=.5cm of loranode0] {Companion App};
\node[rtblock] (companion1) [below=.5cm of loranode1] {Companion App};
\draw [<->] (loranode0) --  node [right,align=center,minimum width=3.5cm] {serial interface} (companion0);
\draw [<->] (loranode1) --(companion1);

\node[rtblock] (lora0) [below=.5cm of companion0] {LoRa Transceiver};
\node[rtblock] (lora1) [below=.5cm of companion1] {LoRa Transceiver};
\draw[<->] (companion0) --  node [right,align=center,minimum width=3.5cm] {SPI} (lora0);
\draw[<->] (companion1) -- (lora1);

\draw[decoration={brace,raise=2pt},decorate]  (script.north east) -- node[below,xshift=3pt,rotate=90,pos=.37] {control domain} (loranode1.south east);
\draw[decoration={brace,raise=2pt},decorate]  (companion1.north east) -- node[below,xshift=3pt,rotate=90,pos=0.25] {real-time domain} (lora1.south east);

\draw[decoration={brace,raise=2pt,mirror},decorate]  (repl.north west) -- node[above,xshift=-3pt,rotate=90,pos=.7] {single instance} (controlinterface.south west);
\draw[decoration={brace,raise=2pt,mirror},decorate]  (loranode0.north west) -- node[above,xshift=-3pt,rotate=90] {per field node} (lora0.south west);

\end{tikzpicture}

%% file: inc/attacks.tex
\section{Wormholes in LoRaWAN}
\label{sec:wormholes}

The concept of wormholes in LoRaWAN is different from the typical understanding, as LoRa is only utilized on a single hop between \acp{ed} and  \acp{gw}.
Usually, wormholes exploit a faster out-of-bound connection to relay messages of a multi-hop network, exceeding the network's own forwarding speed.
All LoRaWAN wormholes follow a store-and-forward principle to bridge a single hop.

Since communication of LoRaWAN Class A \acp{ed} is only uplink-induced, bidirectional LoRaWAN wormholes are constrained by the receive windows of a transaction to forward downlink messages.
We use a simple unidirectional wormhole as a foundation to pre\-sent two more sophisticated, downlink-enabled approaches.
We then incorporate them into an attack exploiting the \ac{adr} mechanism of LoRaWAN.

\subsection{Unidirectional Wormhole}

\begin{figure}
	\centering
	\input{img/unidirectional-wormhole}
	\Description{The figure shows a sequence diagram for the end device on the left, the gateway on the right, and the attacker with two nodes in the center. The end device starts transmitting a frame $up_n$, which is intended for the gateway. However, the arrow stops just before reaching the gateway with a label stating that the frame is jammed. Simultaneously, the attacker node which is close to the end device receives the frame. The attacker internally forwards the sniffed frame to her other node which is close to the gateway. From there, the frame is replayed as $up_n'$. The time between the end of the original frame and the start of the replayed frame is labeled $t_processing$. The time it takes to transmit the uplink frame is called $t_uplink$. After a delay of $d_{rx_1}$, the gateway answers with a frame labeled $down_n$. While the arrow for this frame reaches the end device, the figure shows that the end device's receive windows ``rx1'' and ``rx2'' are missed.}
	\caption{Message flow of the unidirectional wormhole}
	\label{fig:unidirectional-wormhole}
\end{figure}
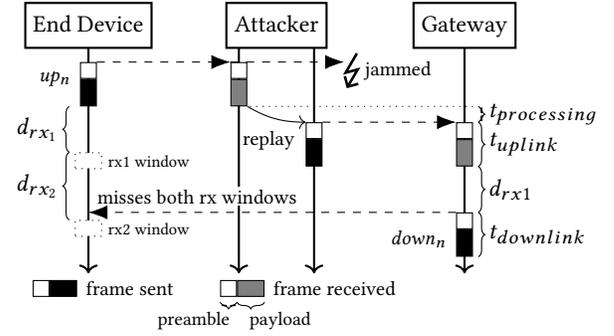

Figure \ref{fig:unidirectional-wormhole} shows the message flow within a unidirectional, uplink-only wormhole:
One of the attacker's transceiver acts as the entry node and is placed near the target \ac{ed}.
Another transceiver serves as an exit node near the \acp{gw}.
A LoRa frame $up_n$ is captured in its entirety by the entry node and forwarded to the wormhole's exit node through an out-of-bound channel.
The exit node then replays the message $up_n'$ for the \ac{gw} to receive.

If the \ac{ed} is within reach of the \ac{gw}, the exit node can be configured to selectively jam the reception of frames while the entry node still receives them.
This approach differs from the LoRa wormhole introduced in \cite{aras2017selective} by the jammer's trigger.
We discovered that listening on the \emph{exit node} and quickly switching it to jamming is significantly faster than triggering the jammer by the sniffer via network.
This allows selective jamming based on the \texttt{DevAddr} even for SF7 and still receiving the messages at the entry node, if the jammer's signal is at least \SI{6}{\dB} weaker at the source \cite{goursaud2015dedicated}.

A wormhole cannot alter the LoRaWAN message content, as it is protected by a \ac{mic}.
However, the attacker can modify metadata of the message, in particular the timing, the location, \ac{snr}, and \ac{rssi} values.
If the device and network under attack both conform to the LoRaWAN~1.1 specification, an attacker has to replay uplink messages on the same channel and \ac{dr}, as those parameters take part in the \ac{mic} calculation in the updated specification \cite[Section 4.4.2]{lora2017specs11}.
That was not the case in LoRaWAN~1.0.x \cite[Section 4.4]{lora2018specs103}, which allows replay attacks with different transmission parameters.

\subsection{RX2 Wormhole}

The attacker cannot replay the downlink message in the rx1 window of a transaction.
Figure \ref{fig:unidirectional-wormhole} illustrates how the rx1 window starts at a fixed delay of $d_{rx_1}$ measured from the end of the uplink message $up_n$.
So when receiving the replayed message $up_n'$, the \ac{ns} will respond not earlier than $d_{rx_1}$ after that.

The transmission parameters of LoRaWAN downlink messages are not protected by the message's \ac{mic}, even in LoRaWAN~1.1 \cite[Section 4.4.1]{lora2017specs11}.
Therefore, an attacker may aim for the second receive window to forward the downlink message.
The additional delay can be exploited to circumvent the timing constraints.

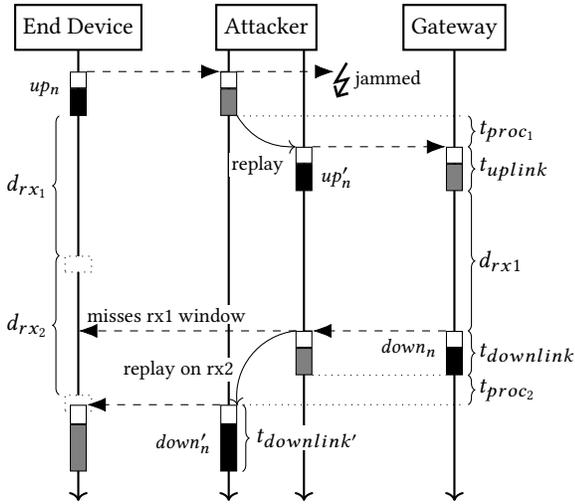
\begin{figure}
	\centering
	\input{img/rx2-wormhole}
	\Description{A sequence diagram showing the actions of the end device on the left, the gateway on the right, and the attacker with two nodes in the center. The end device transmits an uplink frame called $up_n$ of duration $t_{uplink}$. The frame is intended for the gateway, but the arrow does not reach it. Instead, the frame is labeled as jammed and only received by the attacker on the node near the end device. The attacker replays this frame after a period called $t_{processing_1}$. The replayed frame is called $up_n'$. After a period called $d_rx_1$, the gateway answers with a downlink frame called $down_n$. The frame as a duration of duration of $t_{downlink}$ $d_{rx2}$ and is received by the attacker on the node near the gateway. The arrow for this frame also reaches the end device, but misses the ``rx1'' receive window. The frame then is replayed by the attacker after an additional period called $t_{proc_2}$ on the node near the end device, now called $down_n'$. The transmission now takes longer than before, this time is labeled as $t_downlink'$. The frame reaches the end device in its ``rx2'' receive window.}
	\caption{Message flow of the rx2 wormhole}
	\label{fig:rx2-wormhole}
\end{figure}

Figure \ref{fig:rx2-wormhole} illustrates the principle of the \emph{rx2 wormhole}.
Downlink messages are scheduled to match the rx2 window.
The diagram also illustrates the remaining timing constraints of such a wormhole.
The duration of the replayed uplink $t_{uplink}$, of the original downlink $t_{downlink}$, and additional time for processing, $t_{proc_1}$ and $t_{proc_2}$, all together must not exceed the duration of the time between both receive windows:

\begin{equation}
\begin{aligned}
	t_{proc_1} + t_{uplink} + d_{rx_1} + t_{downlink} + t_{proc_2} & \leq d_{rx_1} + d_{rx_2} \\
	t_{proc_1} + t_{uplink} + t_{downlink} + t_{proc_2} & \leq \SI{1}{\second} \hspace{6pt} (= d_{rx_2})
\end{aligned}
\end{equation}

\begin{figure}[t]
	\centering
	\input{img/frame-transmission-time}
	\Description{The chart shows a line for each data rate for payload lengths from 1 to 30 bytes. The LoRaWAN minimum payload length of 12 bytes is marked as well as a limit of 500ms for the frame time. Data rates 2 to 6 are below these limitations while data rates 1 and 0 clearly exceed them.}
	\caption[LoRa frame transmission time in relation to data rate (EU868) and payload length]{LoRa frame transmission time in relation to \ac{dr} (\texttt{EU868}) and payload length. LoRaWAN uplink configuration with header and payload CRC on physical layer}
	\label{fig:frame-transmission-time}
\end{figure}
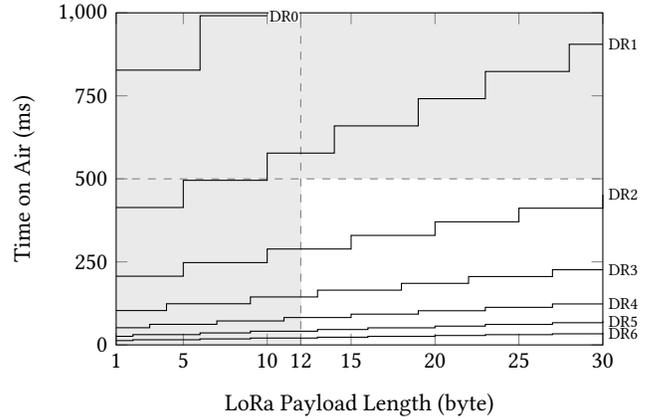

Assuming a negligible processing time and equal frame lengths for uplink and downlink, we get an upper boundary for the transmission time of \SI{500}{\milli\second} per frame.
A LoRa frame containing the LoRaWAN data has a length of at least \SI{12}{bytes}.
Figure \ref{fig:frame-transmission-time} puts both of these limitations in relation to the \acp{dr} for the \texttt{EU868} region.

It is evident that the rx2 wormhole breaks the timing constraints for DR0 and DR1. For DR2 and DR3, it can be used if small or medium-sized LoRaWAN payloads are expected.
These downsides are compensated by the advantage of keeping both frames in the same transaction.
This is in particular important for confirmed uplinks, as the \ac{mic} of the corresponding downlink is only valid during the same transaction in LoRaWAN~1.1 \cite[Section 4.4.1]{lora2017specs11} as a response to the ACK spoofing attack on LoRaWAN~1.0 \cite{yang2017lorawan,donmez2018security}.

\subsection{Downlink-Delayed Wormhole}

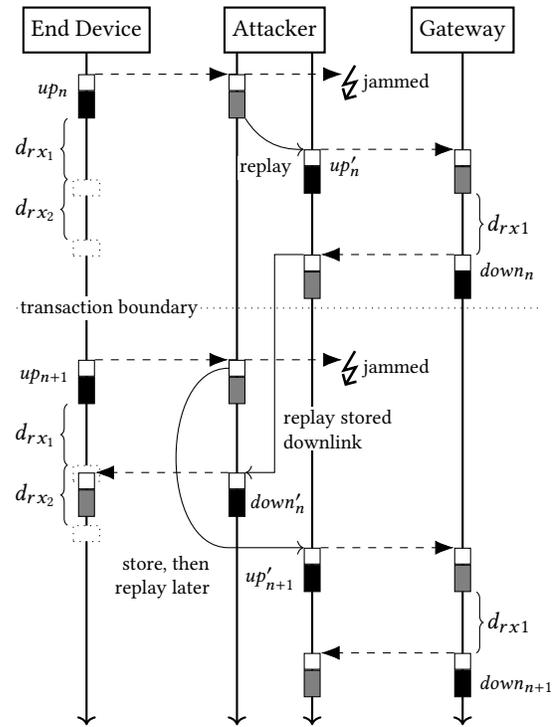
\begin{figure}
	\centering
	\input{img/downlink-delayed-wormhole}
	\Description{The figure shows a sequence diagram of the end device on the left, the gateway on the right, and the attacker with two nodes in the center. The end device starts by sending frame $up_n$, which is intended for the gateway. The arrow does not reach the gateway, but is labeled with ``jammed''. The attacker sniffs the frame with the node close to the end device. Then, the attacker replays the frame as $up_n'$ to the gateway. After a period called $d_{rx_1}$, the gateway answers with a frame called $down_n$, which is received by the attacker's node close to the gateway. Meanwhile, both receive windows ``rx1'' and ``rx2'' have passed for the end device. The diagram shows the transaction boundary at this point. The end device then again transmits an uplink frame $up_{n+1}$, which is jammed for the gateway and received by the attacker like before. Now, the attacker waits for the end device to open the receive window called ``rx1''. This happens at a delay of $d_{rx_1}$ after the transmission of $up_{n+1}$ is finished. She then replays the previously captured $down_n$. After that, the attacker replays the stored uplink $up_{n+1}$ to the gateway, labeled as $up_{n+1}'$. After a delay of $d_{rx_1}$ the gateway responds with a downlink $down_{n+1}$, which is captured by the attacker.}
	\caption{Message flow of the downlink-delayed wormhole}
	\label{fig:downlink-delayed-wormhole}
\end{figure}

By expanding the wormhole concept over two consecutive transactions, an attacker achieves the ability to target low \acp{dr} at the cost of not being able to handle confirmed uplink messages in LoRaWAN 1.1.
We call this approach the \emph{downlink-delayed wormhole}.
Figure \ref{fig:downlink-delayed-wormhole} shows how the uplink message $up_n$ is intercepted, sniffed, and then replayed as $up_n'$.
If the \ac{ns} responds with a downlink message $down_n$, it is not immediately forwarded but stored by the attacker.

Once the \ac{ed} transmits the next uplink message $up_{n+1}$, it is again sniffed and jammed.
If a pending downlink message has been stored by the attacker, the priority is to forward it back to the \ac{ed} through the entry node.
As the entire message is already available, the attacker can aim 
for the rx1 window.
When the downlink message has been delivered, the attacker eventually forwards the stored uplink message $up_{n+1}$ through the exit node to the \ac{ns}, which may return the next pending downlink $down_{n+1}$.

This approach overcomes the issues with high \acp{dr} at the cost of crossing transaction boundaries.
For multi-channel networks, this also adds complications regarding the downlink \ac{fcnt}.
When staying within the same transaction, a downlink message from the \ac{ns} always contains the current maximum for the \ac{fcnt}.
This is important since the \ac{ed} will only accept messages with a higher \ac{fcnt}.
If an attacker cannot observe all channels at all \acp{dr}, she may miss a transaction in which a higher downlink \ac{fcnt} is processed by the \ac{ed}.
In that case, the stored downlink frame of the attacker becomes outdated, as its \ac{fcnt} prevents it from being accepted.

\subsection{From Wormholes to ADR Spoofing}

The \emph{\ac{adr} spoofing attack} exploits the \ac{adr} mechanism to force the \ac{ed} into using a \ac{tp} and \ac{dr} at which it is unable to communicate with any \ac{gw}.
An attacker can intentionally create such a situation by selectively forwarding messages through a wormhole to manipulate their metadata.
The \ac{ed}'s \ac{rssi} and \ac{snr} then appear as being higher for the \ac{ns}.
Employing these values in the \ac{adr} calculation leads to too optimistic estimates for the target \ac{tp} and \ac{dr}.

Once the \ac{ed} adapted to the new \ac{dr}, the attacker decides which messages to forward to the \ac{ns}.
To keep the \ac{ed} in this state, at least one transaction out of \texttt{ADR\_\allowbreak{}ACK\_\allowbreak{}LIMIT} + \texttt{ADR\_\allowbreak{}ACK\_\allowbreak{}DELAY} must pass the wormhole.
This prevents the \ac{ed} from increasing \ac{tp} and lowering the \ac{dr} because its \texttt{ADR\_\allowbreak{}ACK\_\allowbreak{}CNT} is reset to \num{0}.

From this description, we can identify two phases of the attack:
First, the spoofing phase, in which the attacker intercepts messages on the initial \ac{dr}, and second, the retention phase, in which the \ac{ed} is forced to keep its settings by selective forwarding.

While the decision to enable \ac{adr} is made on the \ac{ed}, the choice of the actual parameters is made on the \ac{ns} and announced in a \texttt{LinkADRReq} MAC command.
Therefore, the attacker has two concerns during the spoofing phase:
First, to forward uplink frames with good reception parameters to trigger a \texttt{LinkADRReq} command with a high \ac{dr}, and second, to forward the downlink message containing this command to the \ac{ed}.
This calls for the deployment of a bidirectional wormhole.
Since the attack is most effective if the \ac{ed} can only communicate on low \acp{dr} by default, the attacker may need to pick the downlink-delayed variant for spoofing.

Once the \ac{ed} has processed the \texttt{LinkADRReq}, it immediately switches to the higher \ac{dr} and confirms the new settings with a \texttt{LinkADRAns} in the next uplink message.
The attacker does not forward this message, as a chance exists that the \texttt{LinkADRReq} will remain unacknowledged in the \ac{ns}'s MAC command queue.
If that is the case, the \ac{ns} itself will push the \ac{ed} back to the higher \ac{dr} without any additional effort, should a link be reestablished.
Since MAC commands are only acknowledged once \cite[Section 5, Note 2]{lora2017specs11}, this can create a self-reinforcing situation until the \texttt{LinkADRReq} at the \ac{ns} times out.

To not rely only on the \ac{ns}, the attacker also creates a wormhole on the ``optimized'' \ac{dr}.
For this \ac{dr}, the rx2 wormhole is sufficient, which comes with fewer complications.
We assume an ideal case of no communication between the \ac{ed} and \ac{gw} without the help of the attacker.
The most reliable strategy for the attacker to achieve her goals then would be to selectively forward a transaction if the \texttt{ADRAckReq} flag is set.
The \ac{ns} processes the message and ideally issues a downlink message to reset the \texttt{ADR\_\allowbreak{}ACK\_\allowbreak{}CNT} of the \ac{ed}.
If the \texttt{ADRACKReq} flag is unset in the next transaction, the \ac{ed} has processed the downlink and \texttt{ADR\_\allowbreak{}ACK\_\allowbreak{}CNT} is reset.

The most critical task for the attacker is to ascertain whether and when the \ac{ed} changed the \ac{dr}.
The plain \texttt{LinkADRReq} command can only be observed if it is transmitted piggy-backed in a LoRaWAN~1.0 setup.
In all other situations, MAC commands are encrypted, which only allows inferring the total length of all MAC commands in the message.
Furthermore, observing downlink messages does not guarantee that they are received and processed by the \ac{ed}.
With the given capabilities and a single node, the attacker cannot receive the uplink message containing \texttt{LinkADRAns} after \ac{dr} adaption and still observe the initial channel.
In that case, a probabilistic approach makes sense.
With $n$ being the number of channels on the network and $f_{up}$ being the \ac{ed}'s uplink periodicity, we calculate:

\begin{equation}
t_{timeout} = \frac{1}{f_{up}} * \left\lceil \log_{\frac{n-1}{n}}(0.01) \right\rceil
\end{equation}

After not receiving messages for $t_{timeout}$, an attacker listening to a single channel can assume that the \ac{ed} has switched the \ac{dr} in 99\% of all cases and proceed to the retention phase.
For a network with \num{3} channels, this corresponds to \num{12} uplink messages.

\section{Beacon Spoofing}
\label{sec:beacon-spoofing}

\begin{figure}
	\centering
	\input{img/beacon-spoofing-drift}
	\Description{A diagram with the time as relative offset on the x-axis and the beacon period on the y-axis. For each beacon period, a beacon frame consisting of a preamble and a payload is shown. In period -1, the true beacon frame starts at a time $t_0$. Starting at period 0, spoofed beacons are shown gradually shifted by $\Delta t_{step}$ each period. In the last period, the spoofed beacon is shifted so that the end of the payload is aligned to the start of the true beacon $t_0$. Each spoofed beacon is followed by a ``jammer payload'' of the same length as the frame including the preamble. This means that for the last period, the jammer payload is exactly aligned to the true beacon.}
	\caption{Concept of the beacon drifting attack}
	\label{fig:beacon-spoofing-drift}
\end{figure}
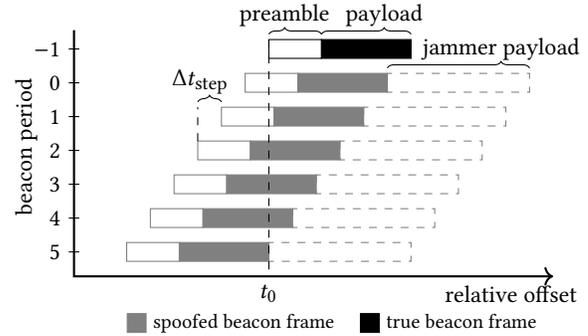

Another attack that creates a \ac{dos} situation by tampering with network parameters is beacon spoofing.
In contrast to \ac{adr} spoofing, 
beacon spoofing affects all Class~B \acp{ed} within reach of the attacker.
The affected devices lose their ability to receive network-induced downlink traffic.

The concept behind this attack has been discussed \cite{vanes2018denial,butun2018analysis}, but to our knowledge, no proof exists showing its practical applicability.
It is based on everyone's ability to create valid Class~B beacon frames, which inherently is required for the coexistence of LoRaWAN networks with different operators.
If an attacker transmits beacon frames with an offset, all \acp{ed} locked to the fake beacon do not open their receive windows on time.

While periodically transmitting frames is a quite feasible task, forcing already locked \acp{ed} to switch to the spoofed beacons requires a bit more effort.
Locked devices open their beacon receive window only just before they expect to receive a beacon.
\acp{ed} searching for beacons utilize \texttt{DeviceTimeReq} commands to reduce the size of their beacon search window.
Targeting only \acp{ed} in the beacon acquisition phase reduces the applicability of the attack drastically.
Therefore, we focus on already locked \acp{ed} only.

By first locking to the legitimate beacon herself, an attacker synchronizes with the network's time without access to GPS.
Then, she starts the beacon drifting as shown in Figure~\ref{fig:beacon-spoofing-drift}.
While staying within the receive window tolerance of the \acp{ed}, she slowly prepones the transmission of the spoofed beacon by $\Delta t_{step}$ each beacon period.
If the step size is chosen small enough, the \acp{ed} shift their beacon receive window together with the spoofed beacon.
Once the accumulated drift corresponds to at least an entire beacon frame length, the attacker can stop shifting.
Using this stop criterion aims at minimizing the collision time with the legitimate beacon.
The beacon frame duration in \texttt{EU868} is \SI{152.58}{\milli\second}.
For our experiments, we add a margin of \num{5} symbols, which corresponds to half a beacon preamble, leading to a total drift of \SI{173.06}{\milli\second}.
\acp{ed} open their receive windows early by the same offset and close them before a downlink transmission starts.

As the beacon frames are sent in implicit header mode without a CRC on the physical layer, the attacker can bring in another measure to increase her success rate.
Bytes added to the end of the beacon frame are discarded by all receivers, as the expected frame length has to be known in this transmission mode.
The attacker can exploit this by adding random data as a jamming payload after the beacon data, which then coexists with the legitimate beacon frame, making it harder for \acp{ed} to stay locked with it.

If the attacker is successful, the \ac{ed} sticks to the spoofed beacon as long as the attacker transmits it, because the \ac{ed} can only detect beacon presence, but not downlink availability.
Once the attacker stops transmitting beacons, the \acp{ed} perform two hours of beaconless operation and eventually return to Class~A.

%% file: img/unidirectional-wormhole.tex
\begin{tikzpicture}[
	y=-1cm,
	node distance=0,
]
\pgfmathsetmacro{\xed}{1}
\pgfmathsetmacro{\xatk}{3.5}
\pgfmathsetmacro{\xatkleft}{\xatk-0.5}
\pgfmathsetmacro{\xatkright}{\xatk+0.5}
\pgfmathsetmacro{\xgw}{6}
\pgfmathsetmacro{\rxdelay}{0.9}
\pgfmathsetmacro{\offsettop}{0.6}
\pgfmathsetmacro{\tprocessing}{0.5}
\pgfmathsetmacro{\symbheight}{0.1}
\pgfmathsetmacro{\frameheight}{0.3}
\pgfmathsetmacro{\framewidth}{0.3}

\pgfmathsetmacro{\ylen}{\offsettop+3*\frameheight+\tprocessing+\rxdelay+0.4}

\draw [->,thick] (\xed, 0)
	node [draw,fill=white] (toped) {{\strut}End Device}
	-- (\xed, \ylen);
\draw [->,thick] (\xatkleft,0) -- (\xatkleft, \ylen);
\draw [->,thick] (\xatkright,0) -- (\xatkright, \ylen);
\node [thick,draw,fill=white] (topatk) at (\xatk, 0) {{\strut}Attacker};
\draw [->,thick] (\xgw, 0)
	node [draw,fill=white] (topgw) {{\strut}Gateway}
	-- (\xgw, \ylen);

\pgfmathsetmacro{\frametop}{\offsettop}
\tikzdrawframe{\xed}{\frametop}{black}{tx1_a_ed}{1}
\node [left=of tx1_a_ed_preamble.south west,annotation] {$up_n$};
\tikzdrawframe{\xatkleft}{\frametop}{gray}{tx1_a_atk}{1}
\node [inner sep=0] (jam) at ({\xatkright+1/4*(\xgw-\xatkright)},\frametop) {\includegraphics[scale=0.5]{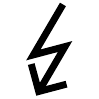}};
\node [annotation] (jam_lbl) [right=-.2 of jam] {jammed};
\draw [txarrow] (tx1_a_ed_preamble.north east) -- (tx1_a_atk_preamble.north west);
\draw [txarrow,shorten >=.1cm] (tx1_a_atk_preamble.north west) -- (jam|-tx1_a_atk_preamble.north);

\tikzdrawwindow{\xed}{\frametop+\frameheight+\rxdelay}{rx1window}
\node [right=of rx1window,scale=0.7] {rx1 window};
\draw[decoration={brace,raise=2pt,mirror},decorate] 
(rx1window.west |- tx1_a_ed_payload.south west) -- node[left,xshift=-3pt] {$d_{rx_1}$} (rx1window.north west);
\tikzdrawwindow{\xed}{\frametop+\frameheight+2*\rxdelay}{rx2window}
\node [right=of rx2window,scale=0.7] {rx2 window};
\draw[decoration={brace,raise=2pt,mirror},decorate] 
(rx1window.north west) -- node[left,xshift=-3pt] {$d_{rx_2}$} (rx2window.north west);

\pgfmathsetmacro{\frametop}{\frametop+\frameheight+\tprocessing}
\tikzdrawframe{\xatkright}{\frametop}{black}{tx1_b_atk}{1}
\node [left=of tx1_b_atk_preamble.south west,annotation] {$up_n'$};
\tikzdrawframe{\xgw}{\frametop}{gray}{tx1_b_gw}{1}
\draw [txarrow] (tx1_b_atk_preamble.45) -- (tx1_b_gw_preamble.135);
\draw (tx1_a_atk_payload.south east) edge[bend right=15,->] node[annotation,below left,pos=0.97] {replay} (tx1_b_atk_preamble.north west);
\draw[decoration={brace,raise=2pt},decorate] 
(tx1_b_gw_preamble.north east|-tx1_a_atk_payload.south east) -- node[right,xshift=3pt] {$t_{processing}$} (tx1_b_gw_preamble.north east);
\draw [dotted] (tx1_b_gw_preamble.north east|-tx1_a_atk_payload.south east) -- (tx1_a_atk_payload.south east);
\draw[decoration={brace,raise=2pt},decorate] 
(tx1_b_gw_preamble.north east) -- node[right,xshift=3pt] {$t_{uplink}$} (tx1_b_gw_payload.south east);

\pgfmathsetmacro{\frametop}{\frametop+\frameheight+\rxdelay}
\tikzdrawframe{\xgw}{\frametop}{black}{rx_a_gw}{1}
\node [below left=of rx_a_gw_preamble.west,annotation] {$down_n$};
\draw [txarrow] (rx_a_gw_preamble.north west) -- (toped|-rx_a_gw_preamble.north) node [annotation,above right,align=left] {misses both rx windows};
\draw[decoration={brace,raise=2pt},decorate] 
(tx1_b_gw_payload.south east) -- node[right,xshift=3pt] {$d_{rx1}$} (rx_a_gw_preamble.north east);
\draw[decoration={brace,raise=2pt},decorate] 
(rx_a_gw_preamble.north east) -- node[right,xshift=3pt] {$t_{downlink}$} (rx_a_gw_payload.south east);

\node (legendtoptmp) at(\xed,\ylen+0.2) {};
\node (legendtop) at(toped.west|-legendtoptmp) {};

\node [right=of legendtop,draw,fill=white,rectangle,minimum width=4*1,minimum height=3] (legend_tx_preamble) {};
\node [draw,fill=black,rectangle,minimum width=10*1,minimum height=3] (legend_tx_payload) [right=of legend_tx_preamble] {};
\node (legend_tx_label) [right=of legend_tx_payload,annotation] {frame sent};

\node [right=.5 of legend_tx_label,draw,fill=white,rectangle,minimum width=4*1,minimum height=3] (legend_rx_preamble) {};
\node [draw,fill=gray,rectangle,minimum width=10*1,minimum height=3] (legend_rx_payload) [right=of legend_rx_preamble] {};
\node [right=of legend_rx_payload,annotation] {frame received};
\draw[decoration={brace,raise=2pt},decorate]
(legend_rx_preamble.south east) -- node[below left,xshift=3pt,yshift=-2pt,scale=0.8] {\strut{}preamble} (legend_rx_preamble.south west);
\draw[decoration={brace,raise=2pt},decorate]
(legend_rx_payload.south east) -- node[below right,xshift=-3pt,yshift=-2pt,scale=0.8] {\strut{}payload} (legend_rx_payload.south west);

\end{tikzpicture}

%% file: img/rx2-wormhole.tex
\begin{tikzpicture}[
	y=-1cm,
	node distance=0,
]
\pgfmathsetmacro{\xed}{1}
\pgfmathsetmacro{\xatk}{3.5}
\pgfmathsetmacro{\xatkleft}{\xatk-0.5}
\pgfmathsetmacro{\xatkright}{\xatk+0.5}
\pgfmathsetmacro{\xgw}{6}
\pgfmathsetmacro{\rxdelay}{1.85}
\pgfmathsetmacro{\offsettop}{0.7}
\pgfmathsetmacro{\tprocessing}{0.4}
\pgfmathsetmacro{\symbheight}{0.1}
\pgfmathsetmacro{\frameheight}{0.6}
\pgfmathsetmacro{\framewidth}{0.3}

\pgfmathsetmacro{\ylen}{\offsettop+4.75*\frameheight+2*\tprocessing+\rxdelay+0.1}

\draw [->,thick] (\xed, 0)
	node [draw,fill=white] (toped) {{\strut}End Device}
	-- (\xed, \ylen);
\draw [->,thick] (\xatkleft,0) -- (\xatkleft, \ylen);
\draw [->,thick] (\xatkright,0) -- (\xatkright, \ylen);
\node [thick,draw,fill=white] (topatk) at (\xatk, 0) {{\strut}Attacker};
\draw [->,thick] (\xgw, 0)
	node [draw,fill=white] (topgw) {{\strut}Gateway}
	-- (\xgw, \ylen);

\pgfmathsetmacro{\frametop}{\offsettop}
\tikzdrawframe{\xed}{\frametop}{black}{tx1_a_ed}{1}
\node [left=of tx1_a_ed_preamble.south west,annotation] {$up_n$};
\tikzdrawframe{\xatkleft}{\frametop}{gray}{tx1_a_atk}{1}
\node [inner sep=0] (jam) at ({\xatkright+1/4*(\xgw-\xatkright)},\frametop) {\includegraphics[scale=0.5]{img/icon-flash.pdf}};
\node [annotation] (jam_lbl) [right=-.2 of jam] {jammed};
\draw [txarrow] (tx1_a_ed_preamble.north east) -- (tx1_a_atk_preamble.north west);
\draw [txarrow,shorten >=.1cm] (tx1_a_atk_preamble.north west) -- (jam|-tx1_a_atk_preamble.north);

\tikzdrawwindow{\xed}{\frametop+\frameheight+\rxdelay}{rx1window}
\draw[decoration={brace,raise=2pt,mirror},decorate] 
(rx1window.west |- tx1_a_ed_payload.south west) -- node[left,xshift=-3pt] {$d_{rx_1}$} (rx1window.north west);
\tikzdrawwindow{\xed}{\frametop+\frameheight+2*\rxdelay}{rx2window}
\draw[decoration={brace,raise=2pt,mirror},decorate] 
(rx1window.north west) -- node[left,xshift=-3pt] {$d_{rx_2}$} (rx2window.north west);

\pgfmathsetmacro{\frametop}{\frametop+\frameheight+\tprocessing}
\tikzdrawframe{\xatkright}{\frametop}{black}{tx1_b_atk}{1}
\node [right=of tx1_b_atk_payload.east,annotation] {$up_n'$};
\tikzdrawframe{\xgw}{\frametop}{gray}{tx1_b_gw}{1}
\draw [txarrow] (tx1_b_atk_preamble.north east) -- (tx1_b_gw_preamble.north west);
\draw (tx1_a_atk_payload.south east) edge[bend right=30,->] node[below left,annotation,pos=0.96] {replay} (tx1_b_atk_preamble.north west);
\draw[decoration={brace,raise=2pt},decorate] 
(tx1_b_gw_preamble.north east|-tx1_a_atk_payload.south east)
	-- node[right,xshift=3pt] {$t_{proc_1}$} (tx1_b_gw_preamble.north east);
\draw [dotted] (tx1_a_atk_payload.south east) -- (tx1_b_gw_preamble.east|-tx1_a_atk_payload.south);
\draw[decoration={brace,raise=2pt},decorate] 
(tx1_b_gw_preamble.north east) -- node[right,xshift=3pt] {$t_{uplink}$} (tx1_b_gw_payload.south east);

\pgfmathsetmacro{\frametop}{\frametop+\frameheight+\rxdelay}
\tikzdrawframe{\xgw}{\frametop}{black}{rx_a_gw}{1}
\node [left=of rx_a_gw_payload.north west,annotation] {$down_n$};
\tikzdrawframe{\xatkright}{\frametop}{gray}{rx_a_atk}{1}
\draw [txarrow] (rx_a_gw_preamble.north west)
	-- (rx_a_atk_preamble.north east);
\draw [txarrow] (rx_a_atk_preamble.north east)
	-- (toped|-rx_a_atk_preamble.north east)
	node [above right,annotation] {misses rx1 window};
\draw[decoration={brace,raise=2pt},decorate] 
(tx1_b_gw_payload.south east) -- node[right,xshift=3pt] {$d_{rx1}$} (rx_a_gw_preamble.north east);
\draw[decoration={brace,raise=2pt},decorate] 
(rx_a_gw_preamble.north east) -- node[right,xshift=3pt] {$t_{downlink}$} (rx_a_gw_payload.south east);

\pgfmathsetmacro{\frametop}{\frametop+\frameheight+\tprocessing}
\tikzdrawframe{\xatkleft}{\frametop}{black}{rx_b_atk}{1.75}
\node [below left=of rx_b_atk_preamble.south west,annotation] {$down_n'$};
\tikzdrawframe{\xed}{\frametop}{gray}{rx_b_ed}{1.75}
\draw (rx_a_atk_preamble.north west) edge[bend right=45,->] node[left,align=right,annotation,pos=0.66] {replay on rx2} (rx_b_atk_preamble.north east);
\draw [txarrow] (rx_b_atk_preamble.north west) -- (rx_b_ed_preamble.north east);
\draw[decoration={brace,raise=2pt},decorate] 
(rx_a_gw_payload.south east) -- node[right,xshift=3pt] {$t_{proc_2}$} (rx_a_gw_payload.east|-rx_b_atk_preamble.north);
\draw [dotted] (rx_a_atk_payload.south east) -- (rx_a_gw_payload.south west);
\draw [dotted] (rx_b_atk_preamble.north east) -- (rx_a_gw_payload.south east|-rx_b_atk_preamble.north);
\draw[decoration={brace,raise=2pt},decorate] 
(rx_b_atk_preamble.north east) -- node[right,xshift=6pt,inner sep=1,fill=white] {$t_{downlink'}$} (rx_b_atk_payload.south east);

\end{tikzpicture}

%% file: img/frame-transmission-time.tex
\begin{tikzpicture}[const plot]
	\def\xmax{30}
	\begin{axis}[
		width=\columnwidth*0.95, 
		height=6cm,
		xmin=1,
		xtick={1,5,10,12,15,20,25,30},
		xmax=\xmax,
		ymin=0,
		ymax=1000,
		ytick={0,250,500,750,1000},
		xtick distance=1,
		xlabel=LoRa Payload Length (byte),
		ylabel=Time on Air (\si{\milli\second}),
		clip=false,
		axis background/.style={
			preaction={
				path picture={
					\fill [lightgray!33!white] (axis cs:1,0) rectangle (axis cs:12,1000);
					\fill [lightgray!33!white] (axis cs:12,500) rectangle (axis cs:\xmax,1000);
					\draw [dashed,color=gray] (axis cs:12,0) -- (axis cs:12,1000);
					\draw [dashed,color=gray] (axis cs:1,500) -- (axis cs:\xmax,500);
				}
			}
		}
	]

	\addplot [solid] table[
		x=payload_length,
		y=DR0,
		mark=none,
		col sep=comma,
		restrict y to domain=0:1000,
		restrict x to domain=1:\xmax,
	] {data/frame-transmission-time.csv}
	node [pos=1.0,fill=white,inner sep=0.3,outer sep=1,right,scale=0.7] {DR0};

	\pgfplotsinvokeforeach{DR1,DR2,DR3,DR4,DR5,DR6} {
		\addplot [solid] table[
			x=payload_length,
			y=#1,
			mark=none,
			col sep=comma,
			restrict y to domain=0:1000,
			restrict x to domain=1:\xmax,
		] {data/frame-transmission-time.csv}
		node [pos=1.0,fill=none,right,scale=0.7] {#1};
	}
	
	\end{axis}
\end{tikzpicture}

%% file: img/downlink-delayed-wormhole.tex
\begin{tikzpicture}[
	y=-1cm,
	node distance=0,
]
\pgfmathsetmacro{\xed}{1}
\pgfmathsetmacro{\xatk}{3.5}
\pgfmathsetmacro{\xatkleft}{\xatk-0.5}
\pgfmathsetmacro{\xatkright}{\xatk+0.5}
\pgfmathsetmacro{\xgw}{6}
\pgfmathsetmacro{\rxdelay}{0.8}
\pgfmathsetmacro{\offsettop}{0.7}
\pgfmathsetmacro{\tprocessing}{0.4}
\pgfmathsetmacro{\symbheight}{0.1}
\pgfmathsetmacro{\frameheight}{0.6}
\pgfmathsetmacro{\framewidth}{0.3}

\pgfmathsetmacro{\ylen}{\offsettop*1.5+\frameheight*7+\rxdelay*4+\tprocessing*2}

\draw [->,thick] (\xed, 0)
	node [draw,fill=white] (toped) {{\strut}End Device}
	-- (\xed, \ylen);
\draw [->,thick] (\xatkleft,0) -- (\xatkleft, \ylen);
\draw [->,thick] (\xatkright,0) -- (\xatkright, \ylen);
\node [thick,draw,fill=white] (topatk) at (\xatk, 0) {{\strut}Attacker};
\draw [->,thick] (\xgw, 0)
	node [draw,fill=white] (topgw) {{\strut}Gateway}
	-- (\xgw, \ylen);

\pgfmathsetmacro{\frametop}{\offsettop}
\tikzdrawframe{\xed}{\frametop}{black}{tx1_a_ed}{1}
\node [left=of tx1_a_ed_preamble.south west,annotation] {$up_n$};
\tikzdrawframe{\xatkleft}{\frametop}{gray}{tx1_a_atk}{1}
\node [inner sep=0] (jam) at ({\xatkright+1/4*(\xgw-\xatkright)},\frametop) {\includegraphics[scale=0.5]{img/icon-flash.pdf}};
\node [annotation] (jam_lbl) [right=-.2 of jam] {jammed};
\draw [txarrow] (tx1_a_ed_preamble.north east) -- (tx1_a_atk_preamble.north west);
\draw [txarrow,shorten >=.1cm] (tx1_a_atk_preamble.north west) -- (jam|-tx1_a_atk_preamble.north);

\tikzdrawwindow{\xed}{\frametop+\frameheight+\rxdelay}{rx1window}
\draw[decoration={brace,raise=2pt,mirror},decorate] 
(rx1window.west |- tx1_a_ed_payload.south west) -- node[left,xshift=-3pt] {$d_{rx_1}$} (rx1window.north west);
\tikzdrawwindow{\xed}{\frametop+\frameheight+2*\rxdelay}{rx2window}
\draw[decoration={brace,raise=2pt,mirror},decorate] 
(rx1window.north west) -- node[left,xshift=-3pt] {$d_{rx_2}$} (rx2window.north west);

\pgfmathsetmacro{\frametop}{\frametop+\frameheight+\tprocessing}
\tikzdrawframe{\xatkright}{\frametop}{black}{tx1_b_atk}{1}
\node [right=of tx1_b_atk_preamble.south east,annotation] {$up_n'$};
\tikzdrawframe{\xgw}{\frametop}{gray}{tx1_b_gw}{1}
\draw [txarrow] (tx1_b_atk_preamble.45) -- (tx1_b_gw_preamble.135);
\draw (tx1_a_atk_payload.south east) edge[bend right=30,->] node[annotation,below left,pos=0.96] {replay} (tx1_b_atk_preamble.north west);

\pgfmathsetmacro{\frametop}{\frametop+\frameheight+\rxdelay}
\tikzdrawframe{\xgw}{\frametop}{black}{rx_a_gw}{1}
\tikzdrawframe{\xatkright}{\frametop}{gray}{rx_a_atk}{1}
\node [right=of rx_a_gw_preamble.south east,annotation] {$down_n$};
\draw [txarrow] (rx_a_gw_preamble.north west) -- (rx_a_atk_preamble.north east);
\draw[decoration={brace,raise=2pt},decorate] 
(tx1_b_gw_payload.south east) -- node[right,xshift=3pt] {$d_{rx1}$} (rx_a_gw_preamble.north east);

\pgfmathsetmacro{\frametop}{\frametop+\frameheight}
\draw [dotted] (\xgw+1, \frametop) --
	node [right,annotation,pos=1] {transaction boundary}
	(0, \frametop);

\pgfmathsetmacro{\frametop}{\frametop+\rxdelay}
\tikzdrawframe{\xed}{\frametop}{black}{tx1_a_ed}{1}
\node [left=of tx1_a_ed_preamble.south west,annotation] {$up_{n+1}$};
\tikzdrawframe{\xatkleft}{\frametop}{gray}{tx1_a_atk}{1}
\node [inner sep=0] (jam2) at ({\xatkright+1/4*(\xgw-\xatkright)},\frametop) {\includegraphics[scale=0.5]{img/icon-flash.pdf}};
\node [annotation] (jam2_lbl) [right=-.2 of jam2] {jammed};
\draw [txarrow] (tx1_a_ed_preamble.north east) -- (tx1_a_atk_preamble.north west);
\draw [txarrow,shorten >=.1cm] (tx1_a_atk_preamble.north west) -- (jam2|-tx1_a_atk_preamble.north);

\tikzdrawwindow{\xed}{\frametop+\frameheight+\rxdelay}{rx1windowb}
\draw[decoration={brace,raise=2pt,mirror},decorate] 
(rx1window.west |- tx1_a_ed_payload.south west) -- node[left,xshift=-3pt] {$d_{rx_1}$} (rx1windowb.north west);
\tikzdrawwindow{\xed}{\frametop+\frameheight+2*\rxdelay}{rx2windowb}
\draw[decoration={brace,raise=2pt,mirror},decorate] 
(rx1windowb.north west) -- node[left,xshift=-3pt] {$d_{rx_2}$} (rx2windowb.north west);

\pgfmathsetmacro{\frametop}{\frametop+\frameheight+\rxdelay+0.1}
\tikzdrawframe{\xatkleft}{\frametop}{black}{rx_b_atk}{1}
\node [right=-0.05 of rx_b_atk_payload.east,annotation] {$down_n'$};
\tikzdrawframe{\xed}{\frametop}{gray}{rx_b_ed}{1}
\draw (rx_a_atk_preamble.north west) -- (topatk|-rx_a_atk_preamble.north west) [->] |- node[right,align=left,annotation,pos=0.4] {replay stored\\downlink} (rx_b_atk_preamble.north east);
\draw [txarrow] (rx_b_atk_preamble.north west) -- (rx_b_ed_preamble.north east);

\pgfmathsetmacro{\frametop}{\frametop+\frameheight+\tprocessing}
\tikzdrawframe{\xatkright}{\frametop}{black}{tx2_b_atk}{1}
\node [left=of tx2_b_atk_payload.west,annotation] {$up_{n+1}'$};
\tikzdrawframe{\xgw}{\frametop}{gray}{tx2_b_gw}{1}
\draw [txarrow] (tx2_b_atk_preamble.north east) -- (tx2_b_gw_preamble.north west);
\draw (tx1_a_atk_preamble.west|-tx2_b_atk_preamble.north west) edge[bend left=90] node[below left,align=right,annotation,pos=0.05] {store, then\\replay later} (tx1_a_atk_preamble.west) [->] --  (tx2_b_atk_preamble.north west);

\pgfmathsetmacro{\frametop}{\frametop+\frameheight+\rxdelay}
\tikzdrawframe{\xgw}{\frametop}{black}{rx2_a_gw}{1}
\node [right=of rx2_a_gw_payload.east,annotation] {$down_{n+1}$};
\tikzdrawframe{\xatkright}{\frametop}{gray}{rx2_a_atk}{1}
\draw [txarrow] (rx2_a_gw_preamble.north west) -- (rx2_a_atk_preamble.north east);
\draw[decoration={brace,raise=2pt},decorate] 
(tx2_b_gw_payload.south east) -- node[right,xshift=3pt] {$d_{rx1}$} (rx2_a_gw_preamble.north east);

\end{tikzpicture}

%% file: img/beacon-spoofing-drift.tex
\begin{tikzpicture}
	\pgfmathsetmacro{\stepcount}{6}
	\pgfmathsetmacro{\margintop}{0.3}
	\pgfmathsetmacro{\bcnheight}{0.25}
	\pgfmathsetmacro{\bcngap}{0.2}
	\pgfmathsetmacro{\symbwidth}{0.07}
	\pgfmathsetmacro{\ymargin}{10*\symbwidth}
	\pgfmathsetmacro{\lpreamble}{10*\symbwidth}
	\pgfmathsetmacro{\lpayload}{17*\symbwidth}
	\pgfmathsetmacro{\lbeacon}{\lpreamble+\lpayload}
	\pgfmathsetmacro{\ljammer}{\lbeacon}
	\pgfmathsetmacro{\stepsize}{\lbeacon/\stepcount}
	\pgfmathsetmacro{\xmin}{-\lbeacon-\ymargin}
	\pgfmathsetmacro{\totalheight}{\margintop+(\bcngap+\bcnheight)*(\stepcount+1)}
	
	\draw [->,thick] (\xmin,\totalheight-\margintop+\bcnheight/2) coordinate (yaxis)
	|- (\lbeacon+\ljammer,0) node (xaxis) [below,xshift=-16pt] {relative offset};
	\node at (\xmin,\totalheight/2) [rotate=90,yshift=18pt] {beacon period};
	
	\pgfmathsetmacro{\bcntop}{\totalheight-\margintop}
	\draw[color=black] (0,\bcntop) rectangle
		(\lpreamble,\bcntop-\bcnheight);
	\draw[decoration={brace,raise=2pt,aspect=0.3},decorate] 
		(0,\bcntop) -- node[above,pos=0.3,yshift=2pt] {preamble} (\lpreamble,\bcntop);
	\draw[fill=black,color=black] (\lpreamble,\bcntop) rectangle
		(\lbeacon,\bcntop-\bcnheight);
	\draw[decoration={brace,raise=2pt,aspect=0.7},decorate]
		(\lpreamble,\bcntop) -- node[above,pos=0.7,yshift=2pt] {payload} (\lbeacon,\bcntop);
	\draw ([xshift=2pt]\xmin,\bcntop-\bcnheight/2) --
		([xshift=-2pt]\xmin,\bcntop-\bcnheight/2) node[left] {$-1$};
	
	\foreach \s in {0,...,5} {
		\pgfmathsetmacro{\bcntop}{\totalheight-\margintop-(\s+1)*\bcnheight-(\s+1)*\bcngap}
		\pgfmathsetmacro{\bcnstart}{0-\stepsize*(\s+1)}
		\draw[color=darkgray] (\bcnstart,\bcntop) coordinate (bcnstart\s)
			rectangle (\bcnstart+\lpreamble,\bcntop-\bcnheight);
		\draw[fill=darkgray,color=darkgray] (\bcnstart+\lpreamble,\bcntop)
			rectangle (\bcnstart+\lbeacon,\bcntop-\bcnheight);
		\draw[dashed,color=darkgray] (\bcnstart+\lbeacon,\bcntop) coordinate (bcnend\s)
			rectangle (\bcnstart+\lbeacon+\ljammer,\bcntop-\bcnheight);
		\coordinate (bcnjamend\s) at (\bcnstart+\lbeacon+\ljammer,\bcntop);
		\draw ([xshift=2pt]\xmin,\bcntop-\bcnheight/2) --
			([xshift=-2pt]\xmin,\bcntop-\bcnheight/2) node[left] {$\s$};
	}

	\draw[decoration={brace,raise=2pt,aspect=0.8},decorate] 
	(bcnend0) -- node[above,pos=0.8,yshift=2pt] {jammer payload} (bcnjamend0);

	\draw[dashed] (bcnstart2) -- (bcnstart2 |- bcnstart1);
	\draw[decoration={brace,raise=2pt,aspect=0.5},decorate]
	(bcnstart2 |- bcnstart1) -- node [above,pos=0.1,yshift=2pt] {$\Delta t_\text{step}$} (bcnstart1);

	\draw[dashed] (0,\totalheight-\margintop) --
		(0,0) coordinate (t_beacon) node[below] {$t_0$};

	\node (truelbl) [below=-0.1 of xaxis,scale=0.85,xshift=-22pt] {{\strut}true beacon frame};
	\node (trueicon) [left=0 of truelbl,fill=black,draw=black,minimum height=.25cm,minimum width=.25cm,inner sep=0, outer sep=0] {};
	\node (atklbl) [left=0.25 of trueicon,scale=0.85] {{\strut}spoofed beacon frame};
	\node (atkicon) [left=0 of atklbl,fill=gray,draw=gray,minimum height=.25cm,minimum width=.25cm,inner sep=0, outer sep=0] {};
\end{tikzpicture}

%% file: inc/setup.tex
\section{Experimental Setup}
\label{sec:setup}

An experimental evaluation of the two attacks requires bringing the \frameworkname{} nodes together with LoRaWAN network entities in a testbed.
This section summarizes the network configuration and topology used for our experiments.

\subsection{Network Configuration}

As the specification does not cover all implementation-specific details for the operation of a network, running an experimental evaluation of LoRaWAN always yields results in the context of the selected software.
All experiments in this study use ChirpStack~3.6.

LoRaWAN libraries for \acp{ed} come in much greater variety than software for network infrastructure, with a reference implementation being available.
We deploy its LoRaWAN~1.1 branch\footnote{We use commit \texttt{92e37147} of the \texttt{feature-5.0.0} branch in \url{https://github.com/Lora-net/LoRaMac-node}.} on an ST~Nucleo~L476 evaluation board with an SX1276MB1xAS LoRa transceiver to create our device under test.
The software already comes with sample applications for Class~A and~B operation.

We configure these applications to use \ac{abp} and extend them with remote-control to allow for automation.
These modifications do not affect the behavior of the LoRaWAN implementation with the only exception of forcing the \texttt{nbTrans} parameter to a constant value of 1.
This decision creates comparable conditions for each trial even if the \ac{ns} decides for a different redundancy configuration.

We limit the \ac{ns} and \ac{ed} to use only the three default channels from the \texttt{EU868} region (cf. ETSI EN-330 200-1, Section 7.2.3 \cite{etsi-en-300-220}), at \SI{868.1}{\mega\hertz}, \SI{868.3}{\mega\hertz} and \SI{868.5}{\mega\hertz}. 
This allows verifying that a single LoRa transceiver is sufficient to run attacks against a multi-channel network.
In Section~\ref{sec:evaluation-adr}, we extrapolate the insights from our experiments to networks with larger channel lists.
On the uplink channels, DR0 to DR5 are enabled, which correspond to \acp{sf} 12 to 7, respectively, all at \SI{125}{\kilo\hertz} channel bandwidth.
For the rx2 downlink window and Class~B downlink transmissions, we use the default channel of \SI{869.525}{\mega\hertz} at DR0.
\texttt{RECEIVE\_\allowbreak{}DELAY1} is set to 1~second.

\subsection{Topology}

\begin{figure}[t]
	\centering
	\input{img/channel-baseline-aggregated-freq}
	\Description{The figure shows the LoRaWAN data rate for the EU868 region on the x-axis and has two y-axes. The left one shows the RSSI value in \si{\dBm}, the right one shows the SNR in dB. For all data rates, the RSSI values vary between \SI{-124}{\dBm} and \SI{-126}{\dBm}. SNR values are at around \SI{-10}{\dB}. For data rate 5, no RSSI and SNR measurements are shown as the channel does not allow these data rates. In addition, each data rate has a mark for the minimum SNR required for demodulation. For data rate 0, it is \SI{-20}{\dB}, for data rate 1, it is \SI{-17.5}{\dB}, for data rate 2, it is \SI{-15}{dB}, for data rate 3, it is \SI{-12.5}{\dB}, for data rate 4, it is \SI{-10}{dB}, and for data rate 5, it is \SI{-7.5}{\dB}. Below the chart, the receive rates for each data rate are presented, which are above 95\% for data rate 0 to 3, at 28\% for data rate 4, and at 0\% at data rate 5.}
	\caption[Channel from ED to GW]{Channel from \ac{ed} to \ac{gw} without attacker at highest \ac{tp}. Required \ac{snr} according to \cite{semtech2016adr} (n=300 per \ac{dr})}
	\label{fig:channel-baseline-aggregated-freq}
\end{figure}
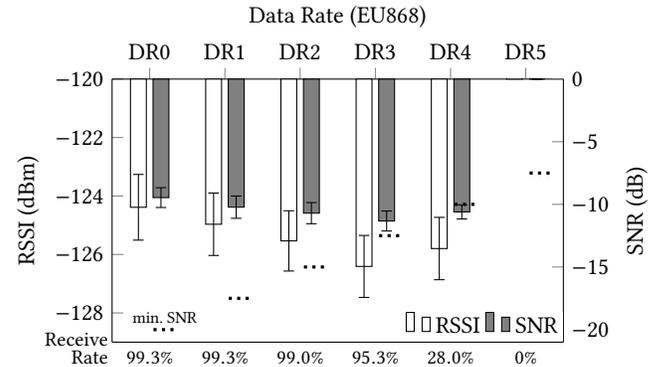

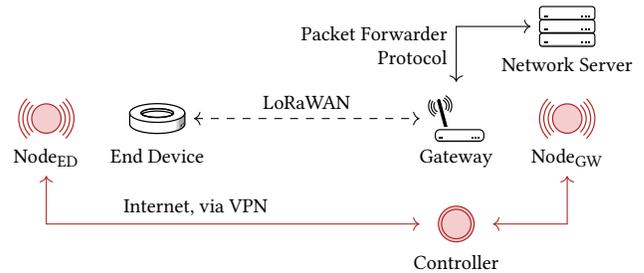
\begin{figure}[t]
	\centering
	\input{img/topology}
	\Description{The figure shows the topology of the experiment. In the center, an end device is connected to a gateway through a LoRaWAN link. The gateway has a connection to a network server using the packet forwarder protocol. The gateway and the end device are each colocated with an attacker node, called Node\textsubscript{GW} and Node\textsubscript{ED}, respectively. Both of them are connected to a controller through an Internet connection using a VPN.}
	\caption{Experiment topology: network under test and \frameworkname{} nodes and controller (red)}
	\label{fig:topology}
\end{figure}

The \ac{ed} and a \ac{gw} are placed at two locations in a way that the channel attenuates the LoRa signal by at least \SI{6}{\dB}, which is a requirement for jamming-based attacks \cite{goursaud2015dedicated}.
This is particularly important if an attacker needs to sniff frames at their source and simultaneously prevent them from reaching their destination.

We also equip the \ac{gw}'s antenna port with an attenuator to fine-tune the \ac{snr} for frames from the \ac{ed} to be just below \SI{-7.5}{\dB}, which is the required \ac{snr} for receiving on DR5 \cite{semtech2016adr}.
This allows us to test the \ac{adr} spoofing attack for a variety of initial \acp{dr}.

Baseline measurements of the channel properties after applying both measures are depicted in Figure~\ref{fig:channel-baseline-aggregated-freq}.
A nearly perfect receive rate for the lower \acp{dr} and its drop for DR4 and DR5 show that the setup is tuned well for the experiments.

With the network being configured, we introduce the \frameworkname{} nodes to the topology, as depicted in Figure~\ref{fig:topology}.
Both LoRaWAN network entities, the \ac{ed} and the \ac{gw}, are each collocated with a single attacker node, called Node\textsubscript{ED} and Node\textsubscript{GW}, respectively.

Each \frameworkname{} node consists of a Pycom LoPy~4 connected to a Raspberry Pi.
It also runs the TPy \emph{LoRa Node} interface for remote-controlling the LoPy~4.
The \frameworkname{} controller, the ChirpStack \ac{ns} API, and the \ac{ed}'s  output are monitored centrally to collect measurements during the experiments.

%% file: img/channel-baseline-aggregated-freq.tex
\begin{tikzpicture}[
		node distance=0,
	]
	\pgfplotsset{
		scale only axis,
		xmin=-0.5,
		xmax=5.5,
		compat=1.3,
		height=3.5cm,
		width=6cm,
	}
	\useasboundingbox (-1.6,-0.35) rectangle (7.4,4.7);

	\begin{axis}[
			ymin=-129,
			ymax=-120,
			xtick distance=1,
			axis x line*=top,
			axis y line*=left,
			xtick={0,1,2,3,4,5},
			xticklabels={DR0, DR1, DR2, DR3, DR4,DR5},
			ylabel=RSSI (\si{\dBm}),
			xlabel=Data Rate (EU868),
			ytick={-120,-122,-124,-126,-128},
			ybar,
			bar width=6pt,
		]

			\addplot+ [
				fill=white,
				draw=black,
				error bars/.cd,
				y dir=both,
				y explicit,
				error bar style={black},
			] table[
				x expr=\thisrow{datarate}-0.15,
				y=rssi_mean,
				y error=rssi_std,
				col sep=comma,
			] {data/channel-baseline-aggregated-freq.csv};
	\end{axis}

	\begin{axis}[
		axis y line*=right,
		axis x line=none,
		ymin=-21,
		ymax=0,
		ytick={0,-5,-10,-15,-20},
		ylabel=SNR (\si{\dB}),
		ybar,
		bar width=6pt,
		legend cell align={left},
		legend columns=3,
		legend style={
			at={(axis cs: 5.5, -21)},
			anchor=south east,
			draw=none,
			inner sep=0,
		},
	]

		\addlegendimage{}\addlegendentry{RSSI}
		\addplot+ [
				fill=gray,
				draw=black,
				error bars/.cd,
				y dir=both,
				y explicit,
				error bar style={black},
			] table[
				x expr=\thisrow{datarate}+0.15,
				y=snr_mean,
				y error=snr_std,
				col sep=comma,
			] {data/channel-baseline-aggregated-freq.csv};
		\addlegendentry{SNR}
        \coordinate (dr0) at (axis cs: 0, 0);
        \coordinate (dr1) at (axis cs: 1, 0);
        \coordinate (dr2) at (axis cs: 2, 0);
        \coordinate (dr3) at (axis cs: 3, 0);
        \coordinate (dr4) at (axis cs: 4, 0);
        \coordinate (dr5) at (axis cs: 5, 0);
        \coordinate (bleft) at (axis cs: -0.5, -21);
        \coordinate (bright) at (axis cs: 5.5, -21);
        \draw [black,dotted,very thick] (axis cs:  0.05, -20.0) -- node [above,scale=0.7] {min. SNR} (axis cs: 0.35, -20.0);
        \draw [black,dotted,very thick] (axis cs:  1.05, -17.5) -- (axis cs: 1.35, -17.5);
        \draw [black,dotted,very thick] (axis cs:  2.05, -15.0) -- (axis cs: 2.35, -15.0);
        \draw [black,dotted,very thick] (axis cs:  3.05, -12.5) -- (axis cs: 3.35, -12.5);
        \draw [black,dotted,very thick] (axis cs:  4.05, -10.0) -- (axis cs: 4.35, -10.0);
        \draw [black,dotted,very thick] (axis cs:  5.05,  -7.5) -- (axis cs: 5.35,  -7.5);

	\end{axis}

	\draw (bleft) -- (bright);
	\node [annotation,below] (firstrr) at (dr0|-bleft) {$99.3\%$};
	\node [annotation,below] at (dr1|-bleft) {$99.3\%$};
	\node [annotation,below] at (dr2|-bleft) {$99.0\%$};
	\node [annotation,below] at (dr3|-bleft) {$95.3\%$};
	\node [annotation,below] at (dr4|-bleft) {$28.0\%$};
	\node [annotation,below] at (dr5|-bleft) {$0\%$};
	\node [left=of firstrr,align=right,annotation,yshift=4,execute at begin node=\setlength{\baselineskip}{8pt}] {Receive\\Rate};

\end{tikzpicture}

%% file: img/topology.tex
\begin{tikzpicture}
	\node (icon_ed) {\includegraphics[scale=0.5]{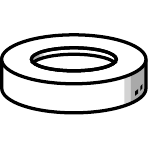}};
	\node [annotation] (text_ed) [below=-0.2 of icon_ed] {End Device};

	\node (icon_gw) [right=3cm of icon_ed] {\includegraphics[scale=0.5]{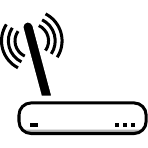}};
	\node [annotation] (text_gw) [below=-0.2 of icon_gw] {Gateway};

	\node (icon_atk_ed) [left=0.5cm of icon_ed] {\includegraphics[scale=0.5]{img/icon-attacker-node.pdf}};
	\node [annotation] (text_atk_ed) [below=-0.2 of icon_atk_ed] {Node\textsubscript{ED}};

	\node (icon_atk_gw) [right=0.5cm of icon_gw] {\includegraphics[scale=0.5]{img/icon-attacker-node.pdf}};
	\node [annotation] (text_atk_gw) [below=-0.2 of icon_atk_gw] {Node\textsubscript{GW}};
	
	\node (icon_atk_ctrl) [below=0.15cm of text_gw] {\includegraphics[scale=0.5]{img/icon-attacker-controller.pdf}};
	\node [annotation] (text_atk_ctrl) [below=-0.2 of icon_atk_ctrl] {Controller};

	\draw [<->,inkscape-firebrick] (text_atk_ed) |- node [black] [annotation,above,pos=0.7] {Internet, via VPN} (icon_atk_ctrl);
	\draw [<->,inkscape-firebrick] (text_atk_gw) |- (icon_atk_ctrl);

	\node (icon_ns) [above=.25cm of icon_atk_gw] {\includegraphics[scale=0.5]{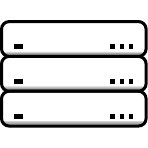}};
	\node [annotation] (text_ns) [below=-0.2 of icon_ns] {Network Server};
	
	\draw [<->,dashed] (icon_ed) -- node [black] [annotation,above] {LoRaWAN} (icon_gw);
	\draw [<->] (icon_gw) |- node [black] [annotation,left,align=right,pos=0.33] {Packet Forwarder\\Protocol} (icon_ns);
\end{tikzpicture}

%% file: inc/evaluation.tex
\section{Results}
\label{sec:evaluation}

For the evaluation, we ran \num{20} trials for each configuration of independent variables.
Before each trial, the \ac{ed} was reactivated on the \ac{ns} using \ac{abp}.
Then the \ac{ed} was reset and the attacker's state was cleared to guarantee independence between trials.

\subsection{ADR Spoofing}
\label{sec:evaluation-adr}

For the wormhole attack on \ac{adr}, we use Node\textsubscript{ED} as the entry node and Node\textsubscript{GW} as the exit node of the wormholes (cf. Figure~\ref{fig:topology}).

\begin{table}[b]
	\caption[Evaluation parameters for the ADR spoofing attack]{Evaluation parameters for the \ac{adr} spoofing attack}
	\label{tab:adr-eval-params}
	\begin{tabular}{ll}
		\toprule
		Parameter & Evaluated Values \\
		\midrule
		spoofing phase: wormhole type & downlink-delayed, rx2 \\
		spoofing phase: data rate & DR0, DR1, DR2, DR3 \\
		uplinks preceding the attack & 1, 10, 20 \\
		\bottomrule
	\end{tabular}
\end{table}

\begin{table}[b]
\caption[Number of transactions from start of attack until the data rate is adjusted to the target DR]{Number of transactions from start of attack until the \ac{dr} is adjusted to the target \ac{dr}}
\label{tab:adr-rampup}
\begin{tabular}{rrr}
	\toprule
	Preceding Uplinks & Number of Transactions (mean, SD) & \multicolumn{1}{c}{$n$} \\
	\midrule
	\begin{filecontents*}{data/adr_rampup_aggregated.csv}
inituplinks,n,requplinksmean,requplinksstd
1,158,7.74,5.0
10,153,7.58,5.79
20,158,7.92,5.57
\end{filecontents*}
\csvreader[
	head to column names,
	late after line = \\
	]{data/adr_rampup_aggregated.csv}{}{
		\num{\inituplinks} & $\requplinksmean \pm \requplinksstd $ & \num{\n}
	}
	\bottomrule
\end{tabular}
\end{table}

We vary the parameters shown in Table~\ref{tab:adr-eval-params}.
We evaluate both wormhole types with all \acp{dr} to see if the actual behavior of the network matches our expectation of the rx2 wormhole being incapable of handling DR0 and DR1 traffic.
The LoRaWAN application data length is set to a single byte to keep it well in the acceptable range for DR2 and DR3.
Changing the number of preceding uplink messages between reset of the \ac{ed} and start of the attack fills the server-side time series of \ac{snr} measurements at different levels.

\subsubsection{Wormhole Type and Data Rate}

\begin{figure}[t]
	\input{img/adr-trig}
 	\Description{The figure shows 8 stacked bars, on for each combination of data rates 0 to 3 with either the downlink-delayed wormhole or the rx2 wormhole. The bars show the percentage for the LinkADRReq being present in a wormhole frame, in an other frame or in no frame, which means that the attack failed. For the downlink-delayed wormhole, the LinkADRReqs in wormhole frames are around a third of all trials. For the rx2 wormhole, data rate 0 and 1 show nearly all cases with ``other frame'', for data rates 2 and 3, nearly all cases show ``wormhole frame''. Lost LinkADRReqs are negligible, only a few cases occur for the rx2 wormhole at data rate 0 and 1.}
	\caption[Trigger for the ED to switch its data rate]{Trigger for the \ac{ed} to switch its \ac{dr} (n=60)}
	\label{fig:adr-trig}
\end{figure}
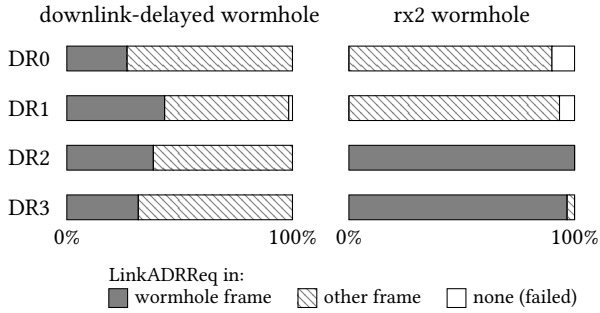

First, we look upon the success of the attacker in the spoofing phase to verify the attack's effectiveness.
We define success as the processing of a \texttt{LinkADRReq} with the target \ac{dr} by the \ac{ed}.
Figure~\ref{fig:adr-trig} shows that this is achieved in most trials, with a few exceptions for the low \acp{dr} and the rx2 wormhole.
In all other cases, the \ac{ed} is forced into the higher \ac{dr}.

Figure~\ref{fig:adr-trig} also depicts how the message containing the critical \texttt{LinkADRReq} command reaches the \ac{ed}.
The results match our expectations about the different capabilities of the wormhole types.

For the rx2 wormhole, a clear distinction exists between \acp{dr}.
On DR2 and DR3, the attacker could directly forward the message to the \ac{ed}.
Notably, the attack does not fail for the lower \acp{dr}.
Frames with high \ac{snr} values still reach the \ac{ns}, and even though the downlink traffic cannot be sent back by the wormhole, the \ac{ns} adds a \texttt{LinkADRReq} to the \ac{ed}'s MAC command queue.
From there, it is delivered on a frequency not observed by the attacker.

\begin{figure}[t]
	\input{img/timing-rx2}
 	\Description{The figure shows a timing diagram for data rates 0, 1, 2, 3, and 5. On the x-axis shows a time period from \SI{-200}{\milli\second} to \SI{2200}{\milli\second}. \SI{0}{\milli\second} are aligned to $t_0$, the time at which the transmission of the original uplink frame $up_n$ is completed. At \SI{1000}{\milli\second}, the ``rx1'' window opens, at \SI{2000}{\milli\second}, the ``rx2'' window opens. For each data rate, the diagram shows the start and duration of the transmission of the replayed frame $up_n'$, the network reply $down_n$, and the replay of the reply $down_n'$. Transmission of $up_n'$ starts at around \SI{150}{\milli\second} for all data rates. The frame transmission time doubles for each data rate, from \SI{46}{\milli\second} for data rate 5 to \SI{1155}{\milli\second} for data rate 0. For data rates 5, 3, and 2, the downlink frame ends before the rx2 window starts with a margin of \SI{790}{\milli\second}, \SI{520}{\milli\second}, and \SI{240}{\milli\second}, respectively. This leaves enough time for replaying a frame in rx2. For data rates 1 and 0, this is not the case, and no replay happens. For data rate 1, the downlink frame is still in transmission during the rx2 windows, for data rate 0, the transmission has not even started.}
	\caption{Measured timing of the rx2 wormhole}
	\label{fig:timing-rx2}
\end{figure}
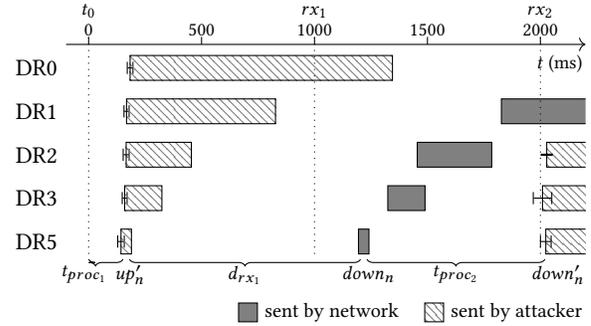

Figure~\ref{fig:timing-rx2} shows the actual timing of the rx2 wormhole during the attack, also including measurements for DR5 from the retention phase.
The attacker is incapable of handling DR1 and DR0 as the downlink frame is not complete when the rx2 window opens.
All other \acp{dr} leave a sufficiently large margin for processing.

For the downlink-delayed wormhole, the results are different, with roughly a third of the messages reaching the \ac{ed} through the wormhole for all \acp{dr}.
The reason for this ratio is the network's channel list's size.
For forwarding, the downlink-delayed wormhole needs two consecutive transactions on the same frequency.
For our three-channel-network, this happens at $33.3\%$ after observing the first transaction, if transactions are treated independently.
So it is reasonable to assume that a full eight-channel-network can be attacked as well with an increase in time to success.

\subsubsection{ADR Algorithm State}

In the next step, we verify that the previous state of the \ac{adr} algorithm on the \ac{ns} does not affect the attack.
Therefore, we vary the number of uplinks preceding the attack and measure the number of transactions before the end device is reconfigured through \ac{adr}.
The result is shown in Table~\ref{tab:adr-rampup}.

As expected, the fill level of the \ac{snr} table on the \ac{ns} did not affect the \ac{adr} decision.
The reason for this is the usage of the max operator instead of averaging over the collected values.

\subsubsection{Retaining the Device}

Once the \ac{ed} processes the \texttt{LinkADRReq}, the attacker proceeds to the retention phase.
The initial \ac{dr}, wormhole type, and \ac{adr} algorithm state do not affect success in retaining, so we aggregate results from all trials reaching this phase.

We first evaluate the overall success of the attacker in this phase, which we define by successfully resetting the \texttt{ADR\_\allowbreak{}ACK\_\allowbreak{}CNT} of the \ac{ed} and by a low ratio of successful uplinks.
In all 469 trials that made it into the retention phase, the attacker was able to retain the \ac{ed} on its high \ac{dr}.
For $95\%$ of trials, the uplink success rate dropped below $2.88\%$, for $99\%$, it was still below $3.11\%$.
This value cannot reach $0\%$ while remaining successful in keeping the \ac{ed} in its state, as \texttt{ADRAckReq} flags have to be answered to.
For our \ac{ed} with a low \texttt{ADR\_\allowbreak{}ACK\_\allowbreak{}LIMIT} of \num{32}, nearly all of the uplink messages are intentional passes through the wormhole.

\begin{figure}
	\input{img/downlink-availability}
	\Description{The figure is a line chart with the beacon period relative to the start of the attack on the x-axis and the downlink availability as percentage on the y-axis. There chart contains for plots for all values of $\Delta t_{step}$: 1, 2, 3, 4, 6, and 8 symbol lengths. At period 0, a vertical line in the chart states that the spoofing starts. The lines for 4, 6, and 8 symbols plateau between 80\% and 100\% for the whole duration of the attack. For 3 symbols, the downlink availability drops to 0\% after 2 periods, for 2 symbols, it drops after 3 periods, and for 1 symbol, it drops after 9 periods.}
	\caption{Downlink availability during beacon drifting}
	\label{fig:downlink-availability}
\end{figure}
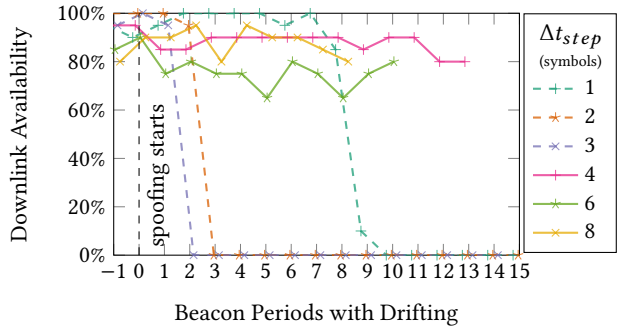

\subsubsection{Countermeasures}

We have seen that tricking the \ac{ed} in using unsuitable \acp{dr} is possible with a high success rate even for LoRaWAN~1.1.
The attack does not have a single enabling vulnerability, but several contributing factors.

The missing relation of messages within a transaction enables the wormhole attacks.
As a reaction to ACK spoofing in LoRaWAN~1.1, the uplink \ac{fcnt} was included in a downlink \ac{mic}, but only if the uplink \texttt{ACK} flag is set.
If that was the case for all Class~A messages, the downlink-delayed wormhole would be prevented.
Also, transmission parameters like \ac{dr} and frequency are only included in the uplink \ac{mic}.
Including them also in downlink \acp{mic} would prevent the rx2 wormhole.
The \texttt{FHDR} field containing the \texttt{ADR} and \texttt{ADRACKReq} flags lacks confidentiality protection.
Protecting \texttt{FHDR} restrains the attacker from identifying frames that need forwarding.

\begin{table}[b]
	\caption[Evaluated values for step size and estimated duration of the beacon drifting attack]{Evaluated values for $\Delta t_{step}$ and estimated duration of the beacon drifting attack}
	\label{tab:beacon-eval-params}
	\begin{tabular}{crcr}
		\toprule
		\multicolumn{2}{c}{$\Delta t_{step}$} & \multicolumn{2}{c}{full beacon length reached after} \\
		symbols & \multicolumn{1}{c}{time} & beacon periods & \multicolumn{1}{c}{time} \\
		\midrule
		\begin{filecontents*}{data/beacon-eval-params.csv}
dtsymb,dtms,reqperiods,reqtime
1,4.096,38,81:04
2,8.192,19,40:32
3,12.288,13,27:44
4,16.384,10,21:20
6,24.576,7,14:56
8,32.768,5,10:40
\end{filecontents*}
\csvreader[
		head to column names,
		late after line = \\
		]{data/beacon-eval-params.csv}{}{\num{\dtsymb} & \SI{\dtms}{\milli\second} & \num{\reqperiods} & \reqtime~min }
		\bottomrule
	\end{tabular}
\end{table}

\ac{adr} algorithms are designed with performance in mind, but not security or robustness.
For example, averaging the \ac{snr} over several messages can prevent abrupt changes from a single data point.
If only uplink messages with plausible \ac{snr} values are directly answered with \texttt{LinkADRReqs} and a strong relation between messages within a transaction is implemented, the success of \ac{adr} spoofing would decrease significantly.

Without these changes, a workaround is to deploy a denser \ac{gw} distribution with fewer \acp{ed} being only in the vicinity of a single \ac{gw} or losing connection on high \acp{dr}.
This is, however, contradictory to the \ac{lpwan}'s paradigm of using a sparse infrastructure.

\subsection{Beacon Spoofing}

Beacon spoofing requires only Node\textsubscript{ED} near the target \ac{ed}.
We vary the drifting step size $\Delta t_{step}$ as shown in Table~\ref{tab:beacon-eval-params}.
For each trial, we wait until at least one downlink message is received in a ping slot after resetting the \ac{ed} to avoid setup problems being falsely attributed to the attacker's success.
We quantify Class~B downlink availability under attack by queuing a downlink message in each beacon period and counting its arrival at the \ac{ed}.
The measurements include the period directly before the attack and three periods after shifting has stopped.

\begin{figure*}
	\input{img/beacon-status}
	\Description{The figure contains six subplots, each for every value of $\Delta t_{step}$. On the x-axis, the beacon periods relative to the attack is shown. The y-axis shows the distribution of the beacon state over all trials in each beacon period. The state can be: valid beacon received, spoofed beacon received, or no beacon received. For 1, 2 and 3 symbol lengths for $\Delta t_{step}$, the state changed from valid beacon received to spoofed beacon received at the moment in which the attack starts. For the remaining time, this state is nearly unchanged, with only occasionally lost beacons in single periods. For the other values, the beacon first gets lost in around 90\% of all trials, and recovers to the true beacon after 5 periods in case of 4 symbols, 3 periods in case of 6 symbols and 2 periods in case of 8 symbols. The spoofed beacon is received only in 5-10\% of all cases.}
	\caption[Beacon status at the ED under attack]{Beacon status at the \ac{ed} under attack}
	\label{fig:beacon-status}
\end{figure*}
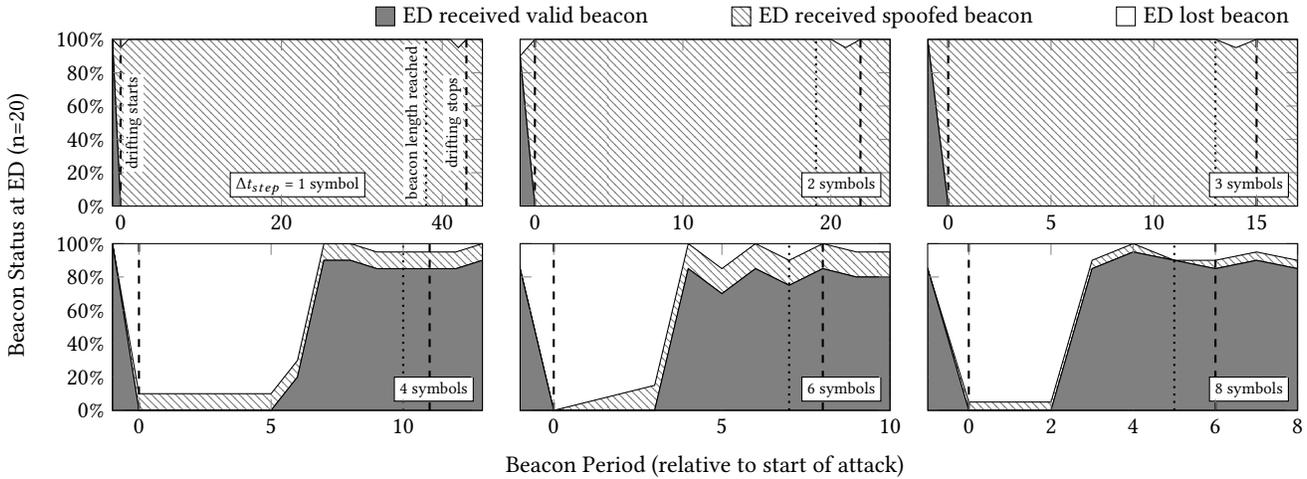

\subsubsection{Impact of Step Size}

Our results in Figure~\ref{fig:downlink-availability} reveal two outcomes:
For $\Delta t_{step}$ of \num{1}, \num{2}, and \num{3} symbols, downlink availability drops significantly after \num{9}, \num{3}, and \num{2} beacon periods, respectively.
For greater values of $\Delta t_{step}$, it remains mostly unaffected at roughly 80\% and above.
We conclude that the attack fails if the beacon is shifted too aggressively, most likely by exceeding the \ac{ed}'s receive window tolerance.

The downlink degrades faster for higher values of $\Delta t_{step}$.
Relating $\Delta t_{step}$ and the beacon period during which the communication breaks down yields a threshold of around 9 symbol lengths.
Once the beacon is shifted further, the timing of the downlink windows between the \ac{ed} and \ac{ns} diverges too far to communicate.

We want to verify that the degrade in downlink quality is indeed caused by the \ac{ed} being locked to the spoofed beacon.
Transmitting a distinct value in the attacker's frames allows distinguishing them from the true ones.
The \texttt{GwInfo} field is well-suited for this purpose, as it does not take part in the calculation of downlink windows.

Figure~\ref{fig:beacon-status} shows the received beacon type over time for each value of $\Delta t_{step}$.
The \ac{ed} may either receive the true, or the spoofed beacon, or no beacon at all.
If a beacon frame cannot be decoded correctly, we count it as lost, since ping slot offsets cannot be calculated without the payload.
Consistent with our previous results, we see that the \ac{ed} is locked to the spoofed beacon if $\Delta t_{step} \leq 3\,\text{symbols}$.
For greater values, the \ac{ed} does not lock to the spoofed beacon but loses track of the true one.
The effect remains for several periods depending on the step size.
For $\Delta t_{step} = 4\,\text{symbols}$, the \ac{ed} recovers after \num{7} periods, while it only takes \num{3} periods if the attacker uses $\Delta t_{step} = 8\,\text{symbols}$.
So, contrary to our expectations, appending jamming payload does not prevent the \ac{ed} from re-locking.

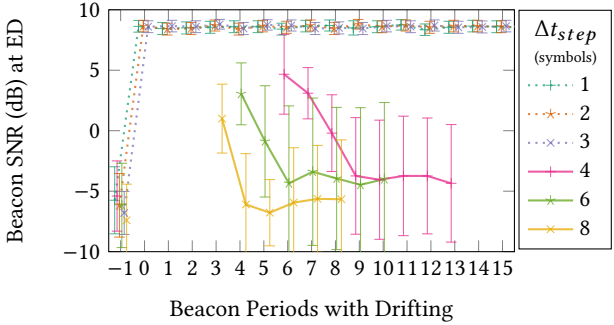
\begin{figure}[t]
	\input{img/beacon-snr}
	\Description{The figure is a line chart with the beacon period relative to the start of the attack on the x-axis, and the beacon SNR measured at the end device in \si{\dB} on the y-axis. On line exists for each value of $\Delta t_{step}$. In the period before the attack starts, all lines start at around \SI{-6}{dB}. For 1, 2, and 3 symbol lengths, the value goes up to around \SI{9}{dB} in period 0 and stays there with a very low deviation. For other values of $\Delta t_{step}$, the lines are discontinued while the beacon is lost, and start again in 3, 4, and 6 for 8, 4, and 6 symbols, respectively. The lines start between 5 and \SI{3}{\dB}, but drop to around \SI{-4}{\dB} within a few periods. The standard deviation is around \SI{5}{dB}, which is bigger than for the other cases by orders of magnitude.}
	\caption[Beacon SNR during attack]{Beacon \ac{snr} during attack. Limited to values for which at least 5 beacons were received}
	\label{fig:beacon-snr}
\end{figure}

To examine the situation further, Figure~\ref{fig:beacon-snr} depicts the beacon \ac{snr} measured at the \ac{ed} during the attack.
For low values of $\Delta t_{step}$, the \ac{snr} increases significantly with the start of the attack and remains high with low variance.
In these cases, the nearby attacker node exploits the capture effect.
The situation is different for the trials with a fast drift.
\ac{snr} levels before the attack starts are comparable.
While the beacon is lost, no values can be measured.
Then, the measurements plateau around a value of \SI{-5}{\dB}, but with a higher variance, which can be ascribed to the presence of the attacker's signal.
Jamming with the extended payload most likely fails since its random symbols do not disturb the receiver's autocorrelation-based detection and synchronization mechanism.

\subsubsection{Countermeasures}

We have shown that an attacker can disrupt the Class~B downlink of \acp{ed} in proximity repeatably in less than ten minutes.
This implies severe consequences for applications relying on a guaranteed downlink latency, especially if no periodic uplink allows for graceful degradation of the service.

With the current requirement for network-spanning beaconing, no effective authentication-based countermeasure is applicable.
Loosening this requirement and returning to network-specific beacons as specified in earlier drafts of LoRaWAN~1.0 \cite[Section 15.1]{lora2016specs102} would allow adding an authentication code to the beacon.
Not transmitting beacons every \SI{128}{\second} based on the GPS epoch, but with an additional network-specific offset in the $[\SI{0}{\second}, \SI{127}{\second}]$ interval, could assure network coexistence.
Since beacons are transmitted with inverse polarity to downlink traffic, additional beacon frames are negligible as a source of interference for the actual downlink traffic on the same channel.
By transmitting the beacon authentication key during \ac{otaa} for specific devices, exposing it in non-volatile device memory is avoided and compromised keys can be replaced.

Without modification of the specification, the attack cannot be prevented.
An \ac{ed} can only try to detect it, for example by observing changes in the beacon's \ac{snr} or a leaping Class~A downlink \ac{fcnt}.
If an attack is suspected, a compensation strategy should be used.
Periodic uplinks assure a minimum level of downlink latency in that case but come at the cost of higher energy consumption.
If the \ac{ns} has evidence that Class~B downlink fails, its only option would be to use a \texttt{ForceRejoinReq} MAC command on the next uplink to force the \ac{ed} to rejoin the network in Class~A \cite[Section 9]{lora2017specs11}.

%% file: img/adr-trig.tex
\begin{tikzpicture}[
	y=-1cm,
]
	\pgfmathsetmacro{\barheight}{0.33}
	\pgfmathsetmacro{\lineheight}{.66}
	\DTLforeach*{trigfile}{\whtype=whtype, \dr=datarate, \direct=direct, \indirect=indirect, \failed=failed}{
		\ifthenelse{\whtype=1}{
			\pgfmathsetmacro{\xoffset}{0}
			\draw (\xoffset,{\dr*\lineheight+\barheight/2})
				node (barddleft\dr) [inner sep=0, outer sep=0] {}
				[fill=gray]
				rectangle ({\xoffset+\direct/20},{\dr*\lineheight-\barheight/2});
			\draw ({\xoffset+\direct/20},{\dr*\lineheight+\barheight/2})
				[pattern=north west lines,pattern color=gray]
				rectangle ({\xoffset+(\direct+\indirect)/20},{\dr*\lineheight-\barheight/2});
			\draw ({\xoffset+(\direct+\indirect)/20},{\dr*\lineheight-\barheight/2})
				[fill=white]
				rectangle ({\xoffset+(\direct+\indirect+\failed)/20},{\dr*\lineheight+\barheight/2}) node (barddright\dr) [inner sep=0, outer sep=0] {};
		}{
			\pgfmathsetmacro{\xoffset}{3.75}
			\draw (\xoffset,{\dr*\lineheight+\barheight/2})
				node (barrx2left\dr) [inner sep=0, outer sep=0] {}
				[fill=gray]
				rectangle ({\xoffset+\direct/20},{\dr*\lineheight-\barheight/2});
			\draw ({\xoffset+\direct/20},{\dr*\lineheight+\barheight/2})
				[pattern=north west lines,pattern color=gray]
				rectangle ({\xoffset+(\direct+\indirect)/20},{\dr*\lineheight-\barheight/2});
			\draw ({\xoffset+(\direct+\indirect)/20},{\dr*\lineheight-\barheight/2})
				[fill=white]
				rectangle ({\xoffset+(\direct+\indirect+\failed)/20},{\dr*\lineheight+\barheight/2}) node (barrx2right\dr) [inner sep=0, outer sep=0] {};
		}
	}
	\draw node at (-.5,\lineheight*0) [align=right] {DR0};
	\draw node at (-.5,\lineheight*1) [align=right] {DR1};
	\draw node at (-.5,\lineheight*2) [align=right] {DR2};
	\draw node at (-.5,\lineheight*3) [align=right] {DR3};
	\draw node at (1.5,-.85*\lineheight) {\strut{}downlink-delayed wormhole};
	\draw node at (5.25,-.85*\lineheight) {\strut{}rx2 wormhole};

	\draw node [below=0 of barddleft3] {0\%};
	\draw node [below=0 of barddright3] {100\%};
	\draw node [below=0 of barrx2left3] {0\%};
	\draw node [below=0 of barrx2right3] {100\%};

	\node (failedlbl) [above=2*\lineheight of barrx2right3,scale=0.85,xshift=-22pt] {{\strut}none (failed)};
	\node (failedicon) [left=0 of failedlbl,fill=white,draw=black,minimum height=.25cm,minimum width=.25cm,inner sep=0, outer sep=0] {};
	\node (indirectlbl) [left=0.25 of failedicon,scale=0.85] {{\strut}other frame};
	\node (indirecticon) [left=0 of indirectlbl,pattern=north west lines,pattern color=gray,draw=black,minimum height=.25cm,minimum width=.25cm,inner sep=0, outer sep=0] {};
	\node (directlbl) [left=0.25 of indirecticon,scale=0.85] {{\strut}wormhole frame};
	\node (directicon) [left=0 of directlbl,fill=gray,draw=black,minimum height=.25cm,minimum width=.25cm,inner sep=0, outer sep=0] {};
	\node (adrlbl) [above right=0 of directicon.north west,scale=0.85,inner sep=0] {{\strut}LinkADRReq in: };

\end{tikzpicture}

%% file: img/timing-rx2.tex
\begin{tikzpicture}[
	y=-1cm,
]
	\pgfmathsetmacro{\barheight}{0.33}
	\pgfmathsetmacro{\scaling}{333}
	\pgfmathsetmacro{\mindata}{5} 
	\pgfmathsetmacro{\yoffset}{-2*\barheight}

	\DTLforeach*{rx2timingfile}{
		\dr=datarate,
		\proca=proc1_mean,
		\procan=proc1_n,
		\procastd=proc1_std,
		\procb=proc2_mean,
		\procbn=proc2_n,
		\procbstd=proc2_std,
		\uplink=uplink_mean,
		\downlink=downlink_mean,
		\downlinkn=downlink_n,
		\rxda=rxdelay1_mean,
		\rxdastd=rxdelay1_std,
		\rxdan=rxdelay1_n,
		\rxdb=rxdelay2_mean,
		\rxdbstd=rxdelay2_std}{

		\pgfmathsetmacro{\xoffset}{0}
		\pgfmathsetmacro{\yoffset}{\yoffset+1.75*\barheight}

		\node at (\xoffset, {\yoffset+0.5*\barheight}) (lblanchor\dr) {};
		\node [left=0.3 of lblanchor\dr,align=right,inner sep=0, outer sep=0] {DR\dr};

		\pgfmathsetmacro{\xoffset}{\xoffset+\proca/\scaling}
		\draw [|-|] (\xoffset-0.5*\procastd/\scaling,\yoffset+0.5*\barheight)
			-- (\xoffset+0.5*\procastd/\scaling,\yoffset+0.5*\barheight);

		\draw [pattern=north west lines,pattern color=gray] (\xoffset, \yoffset) rectangle ({\xoffset+\uplink/\scaling}, \yoffset+\barheight);
		\node at (\xoffset, \yoffset+\barheight) (uplink\dr) {};
		\pgfmathsetmacro{\xoffset}{\xoffset+\uplink/\scaling}
		\node at (\xoffset, \yoffset+\barheight) (uplinkstart\dr) {};
		\node at (\xoffset+\uplink/\scaling/2, \yoffset+\barheight) (uplinkcenter\dr) {};

			\pgfmathsetmacro{\xoffset}{\xoffset+\rxda/\scaling}

		\ifthenelse{\downlinkn>5}{
			\draw [fill=gray] (\xoffset, \yoffset) rectangle ({\xoffset+\downlink/\scaling}, \yoffset+\barheight);
			\node at ({\xoffset+\downlink/\scaling}, \yoffset+\barheight) (downlink\dr) {};
			\node at (\xoffset, \yoffset+\barheight) (downlinkend\dr) {};
			\pgfmathsetmacro{\xoffset}{\xoffset+\downlink/\scaling}
		}{
			\ifthenelse{\dr=1}{
				\draw [fill=gray] (2200/\scaling, \yoffset) -- ({(\proca+\uplink+1000)/\scaling}, \yoffset) |- (2200/\scaling, \yoffset+\barheight);
			}{}
		}

		\ifthenelse{\procbn>\mindata}{
			\pgfmathsetmacro{\xoffset}{\xoffset+\procb/\scaling}
		}{}

		\ifthenelse{\procbn>\mindata}{
			\pgfmathsetmacro{\xoffset}{\rxdb/\scaling}
			\draw [pattern=north west lines,pattern color=gray] (2200/\scaling, \yoffset) -- (\xoffset, \yoffset) |- node (downlinkright\dr) {} (2200/\scaling, \yoffset+\barheight);
			\draw [|-|] (\xoffset-0.5*\rxdbstd/\scaling,\yoffset+0.5*\barheight)
				-- (\xoffset+0.5*\rxdbstd/\scaling,\yoffset+0.5*\barheight);
			\node at (\xoffset, \yoffset+\barheight) (downlinkrep\dr) {};
			\node at (\xoffset+0.1, \yoffset+\barheight) (downlinkrepstart\dr) {};
		}{}

	}

	\draw [->] (-0.3,0-\barheight+0.1) -- node [below=-.05,scale=0.8,pos=.95] {$t$ (\si{\milli\second})} (2200/\scaling, 0-\barheight+0.1);
	\foreach \xval in {0,500,1000,1500,2000}{
		\draw (\xval/\scaling,0-\barheight+0.1) -- node (axis\xval) [above=.025,scale=0.8] {\num{\xval}} (\xval/\scaling,0-\barheight);
	}

	\node [above=0.1 of axis0,align=center,scale=0.8] {$t_0$};
	\node [above=0.1 of axis1000,align=center,scale=0.8] {$rx_1$};
	\node [above=0.1 of axis2000,align=center,scale=0.8] {$rx_2$};
	\draw [dotted] (0, \yoffset+\barheight) -- (0, -\barheight);
	\draw [dotted] (1000/\scaling, \yoffset+\barheight) -- (1000/\scaling, -\barheight);
	\draw [dotted] (2000/\scaling, \yoffset+\barheight) -- (2000/\scaling, -\barheight);

	\node [below=0 of uplinkcenter5,inner sep=0,scale=0.8,xshift=-3] {\strut{}$up_n'$};
	\node [below=0 of downlink5,inner sep=0,scale=0.8] {\strut{}$down_n$};
	\node [below=0 of downlinkrep5,inner sep=0,scale=0.8,xshift=6] {\strut{}$down_n'$};
	\draw[decoration={brace,raise=2pt,mirror,aspect=0.2},decorate]
		(0,\yoffset+\barheight) -- node[below left,yshift=-2pt,xshift=3pt,scale=0.8] {$t_{proc_1}$} (uplinkstart5);
	\draw[decoration={brace,raise=2pt,mirror},decorate]
		(uplink5) -- node[below,yshift=-2pt,scale=0.8] {$d_{rx_1}$} (downlink5);
	\draw[decoration={brace,raise=2pt,mirror},decorate]
		(downlinkend5) -- node[below,yshift=-2pt,scale=0.8] {$t_{proc_2}$} (downlinkrepstart5);
	
	\node (atklbl) [above=1.15 of downlinkright5,scale=0.85,xshift=-13pt] {{\strut}sent by attacker};
	\node (atkicon) [left=0 of atklbl,pattern=north west lines,pattern color=gray,draw=black,minimum height=.25cm,minimum width=.25cm,inner sep=0, outer sep=0] {};
	\node (netlbl) [left=0.25 of atkicon,scale=0.85] {{\strut}sent by network};
	\node (neticon) [left=0 of netlbl,fill=gray,draw=black,draw=black,minimum height=.25cm,minimum width=.25cm,inner sep=0, outer sep=0] {};

\end{tikzpicture}

%% file: img/downlink-availability.tex
\begin{tikzpicture}
	\pgfmathsetmacro{\lineshift}{0.1}
	\tikzset{
		plotstyle/.style={thick,opacity=.75,mark options={thin,solid}},
	}

	\begin{axis}[
		height=4.8cm,
		width=\columnwidth-1.5cm,
		xmin=-1,
		xmax=15, 
		ymin=0,
		ymax=100,
		unbounded coords=jump,
		yticklabel={\pgfmathprintnumber\tick\%},
		xtick distance=1,
		xlabel=Beacon Periods with Drifting,
		ylabel=Downlink Availability,
		cycle list/Dark2,
		legend cell align={left},
		legend columns=1,
		legend style={
			at={(1.01,1)},
			anchor=north west,
		},
		]
		\addlegendimage{empty legend}
		\addlegendentry{\hspace{-.6cm}$\Delta t_{step}$}
		\addlegendimage{empty legend}
		\addlegendentry{\hspace{-.6cm}\scriptsize{(symbols)}}

		\addplot+ [dashed,mark=+,plotstyle]
		table[
			x expr=\thisrow{period}-\lineshift*2.5,
			y expr=\thisrow{availability}*100,
			discard if not={step_size}{1},
			col sep=comma
		] {data/downlink-availability.csv};
		\addlegendentry{1}

		\addplot+ [dashed,mark=star,plotstyle]
		table[
			x expr=\thisrow{period}-\lineshift*0.5,
			y expr=\thisrow{availability}*100,
			discard if not={step_size}{2},
			col sep=comma
		] {data/downlink-availability.csv};
		\addlegendentry{2}

		\addplot+ [dashed,mark=x,plotstyle]
		table[
			x expr=\thisrow{period}+\lineshift*1.5,
			y expr=\thisrow{availability}*100,
			discard if not={step_size}{3},
			col sep=comma
		] {data/downlink-availability.csv};
		\addlegendentry{3}

		\addplot+ [mark=+,plotstyle]
		table[
			x expr=\thisrow{period}-\lineshift*1.5,
			y expr=\thisrow{availability}*100,
			discard if not={step_size}{4},
			col sep=comma
		] {data/downlink-availability.csv};
		\addlegendentry{4}

		\addplot+ [mark=star,plotstyle]
		table[
			x expr=\thisrow{period}+\lineshift*0.5,
			y expr=\thisrow{availability}*100,
			discard if not={step_size}{6},
			col sep=comma
		] {data/downlink-availability.csv};
		\addlegendentry{6}

		\addplot+ [mark=x,plotstyle]
		table[
			x expr=\thisrow{period}+\lineshift*2.5,
			y expr=\thisrow{availability}*100,
			discard if not={step_size}{8},
			col sep=comma
		] {data/downlink-availability.csv};
		\addlegendentry{8}

		\addplot+[
			mark=none,
			black,
			dashed,
			forget plot,
		] coordinates {(0,0) (0,100)}
		node[pos=0.33,right] {\rotatebox{90}{spoofing starts}};

	\end{axis}
\end{tikzpicture}

%% file: img/beacon-status.tex
\begin{tikzpicture}
	\begin{groupplot}[
		group style={
			group size=3 by 2,
			vertical sep=.5cm,
			horizontal sep=.5cm},
		height=3.8cm,
		width=6.5cm]
		\pgfplotsinvokeforeach{1,2,3,4,6,8}{
			\nextgroupplot[
					xmin=-1,
					xmax={1+ceil(173056/4096/#1)+1},
					ymin=0,
					ymax=100,
					yticklabel=\ifthenelse{\equal{1}{#1}\OR\equal{4}{#1}}{\pgfmathprintnumber\tick\%}{},
					xticklabel=\pgfmathprintnumber\tick,
					stack plots=y,
					area style,
					unbounded coords=jump,
			]
				\ifthenelse{\equal{1}{#1}}{
					\node [above] at (axis cs:{(1+ceil(173056/4096/#1)+1)/2-0.5},0) [thin,draw=black,align=center,scale=0.7,fill=white,inner sep=1pt,outer sep=5pt] {\hspace{2pt}\strut{}$\Delta t_{step}$ = 1 symbol\hspace{2pt}};
				}{
					\node [above left] at (axis cs:{(1+ceil(173056/4096/#1)+1},0) [thin,draw=black,align=center,scale=0.7,fill=white,inner sep=1pt,outer sep=5pt] {\hspace{2pt}\strut{}#1 symbols\hspace{2pt}};
				}
				\draw [dotted,thick] (axis cs: {ceil(152576/4096/#1)},0) -- node (bcnline#1) {} (axis cs: {ceil(152576/4096/#1)},100);
				\draw [dashed,thick] (axis cs: {ceil(173056/4096/#1)},0) -- node (stopline#1) {} (axis cs: {ceil(173056/4096/#1)},100);
				\draw [dashed,thick] (axis cs: 0,0) -- node (startline#1) {} (axis cs: 0,100);
				\addplot[
					draw=black,
					fill=gray,
				]
				table[
					mark=none,
					x=period,
					y expr=\thisrow{received}*100,
					discard if not={step_size}{#1},
					col sep=comma,
				]{data/beacon-status.csv}
				\closedcycle;

				\addplot[
					draw=black,
					pattern=north west lines,
					pattern color=gray,
				]
				table[
					mark=none,
					x=period,
					y expr=\thisrow{spoofed}*100,
					discard if not={step_size}{#1},
					col sep=comma,
				]{data/beacon-status.csv}
				\closedcycle;
				
		}
	\end{groupplot}

	\node [right=0 of startline1,scale=0.7,xshift=-5,fill=white,inner sep=1pt,outer sep=2pt] {\rotatebox{90}{drifting starts}};
	\node [left=0 of bcnline1,scale=0.7,xshift=5,fill=white,inner sep=1pt,outer sep=2pt] {\rotatebox{90}{beacon length reached}};
	\node [left=0 of stopline1,scale=0.7,xshift=5,fill=white,inner sep=1pt,outer sep=2pt] {\rotatebox{90}{drifting stops}};

	\path (group c1r1.north west) -- node [above] [rotate=90,yshift=1cm] {Beacon Status at ED (n=20)} (group c1r2.south west);
	\path (group c1r2.south west) -- node [above] [yshift=-1cm] {Beacon Period (relative to start of attack)} (group c3r2.south east);

	\node (lostlbl) [above left=0 of group c3r1.north east] {{\strut}ED lost beacon};
	\node (losticon) [left=0 of lostlbl,draw=black,fill=white,minimum height=.25cm,minimum width=.25cm,inner sep=0, outer sep=0] {};
	\node (spoofedlbl) [left=of losticon] {{\strut}ED received spoofed beacon};
	\node (spoofedicon) [left=0 of spoofedlbl,draw=black,pattern=north west lines,pattern color=gray,minimum height=.25cm,minimum width=.25cm,inner sep=0, outer sep=0] {};
	\node (recvlbl) [left=of spoofedicon] {{\strut}ED received valid beacon};
	\node (recvicon) [left=0 of recvlbl,fill=gray,draw=black,minimum height=.25cm,minimum width=.25cm,inner sep=0, outer sep=0] {};
	
\end{tikzpicture}

%% file: img/beacon-snr.tex
\begin{tikzpicture}
	\pgfmathsetmacro{\lineshift}{0.1}
	\pgfplotsset{
		plotstyle/.style={thick,opacity=.75,mark options={thin,solid},error bars/.cd,y dir=both,y explicit,error bar style={solid}},
	}

	\begin{axis}[
		height=4.8cm,
		width=\columnwidth-1.5cm,
		xmin=-1.5,
		xmax=15.5,
		ymin=-10,
		ymax=10,
		unbounded coords=jump,
		xtick distance=1,
		xlabel=Beacon Periods with Drifting,
		ylabel=Beacon SNR (\si{\dB}) at ED,
		cycle list/Dark2,
		legend cell align={left},
		legend columns=1,
		legend style={
			at={(1.01,1)},
			anchor=north west,
		},
		]
		\addlegendimage{empty legend}
		\addlegendentry{\hspace{-.6cm}$\Delta t_{step}$}
		\addlegendimage{empty legend}
		\addlegendentry{\hspace{-.6cm}\scriptsize{(symbols)}}

		\addplot+ [dotted,mark=+,plotstyle]
		table[
			x expr=\thisrow{period}-\lineshift*2.5,
			y=snr_mean,
			y error=snr_std,
			discard if not={step_size}{1},
			col sep=comma
		] {data/beacon-status.csv};
		\addlegendentry{1}

		\addplot+ [dotted,mark=star,plotstyle]
		table[
			x expr=\thisrow{period}-\lineshift*0.5,
			y=snr_mean,
			y error=snr_std,
			discard if not={step_size}{2},
			col sep=comma
		] {data/beacon-status.csv};
		\addlegendentry{2}

		\addplot+ [dotted,mark=x,plotstyle]
		table[
			x expr=\thisrow{period}+\lineshift*1.5,
			y=snr_mean,
			y error=snr_std,
			discard if not={step_size}{3},
			col sep=comma
		] {data/beacon-status.csv};
		\addlegendentry{3}

		\addplot+ [mark=+,plotstyle]
		table[
			x expr=\thisrow{period}-\lineshift*1.5,
			y=snr_mean,
			y error=snr_std,
			discard if not={step_size}{4},
			col sep=comma
		] {data/beacon-status.csv};
		\addlegendentry{4}

		\addplot+ [mark=star,plotstyle]
		table[
			x expr=\thisrow{period}+\lineshift*0.5,
			y=snr_mean,
			y error=snr_std,
			discard if not={step_size}{6},
			col sep=comma
		] {data/beacon-status.csv};
		\addlegendentry{6}

		\addplot+ [mark=x,plotstyle]
		table[
			x expr=\thisrow{period}+\lineshift*2.5,
			y=snr_mean,
			y error=snr_std,
			discard if not={step_size}{8},
			col sep=comma
		] {data/beacon-status.csv};
		\addlegendentry{8}

	\end{axis}
\end{tikzpicture}

%% file: inc/conclusion.tex
\section{Conclusion}
\label{sec:conclusion}

In this paper, we introduced two attacks affecting the availability of LoRaWAN networks and demonstrated their practical feasibility using \frameworkname{}, our LoRaWAN security evaluation framework.

As the foundation for the novel \ac{adr} spoofing attack, we introduced two concepts of bidirectional wormholes for LoRaWAN which apply even to the latest specification of LoRaWAN~1.1.
Our results show that these wormholes are capable of manipulating frame metadata in LoRaWAN networks.
This enables the \ac{adr} spoofing attack, which allows disrupting the communication for \acp{ed} located at the edge of the network with a high success rate.

We also introduced the concept of beacon drifting as a concrete attack for the vulnerability of missing beacon authentication, which has been mentioned in literature before.
We found that by gradually shifting malicious beacon frames, an attacker may disrupt Class~B downlink communication within an area in less than ten minutes.

We propose countermeasures for both attacks, which require revising the current specification.
Without a change to the specification, the attacks can only be complicated, but not prevented.